\documentclass[11pt]{article}
\usepackage{amsmath,amsfonts,amssymb,latexsym}
\begin{document}
\title{Geometric Complexity Theory II: \\Towards explicit 
obstructions for embeddings among  class varieties}
\author{Dedicated to  Sri Ramakrishna \\ \\ \\ 
Ketan D. Mulmuley \thanks{The work of the first author was
 supported by NSF grant CCR 9800042 and, in part, by Guggenheim Fellowship.}
\\
The University of Chicago and I.I.T., Mumbai \thanks{Visiting faculty member}
\\ Milind Sohoni\\ I.I.T., Mumbai \\ \\  (To appear in SIAM J. Comp.) 
}

\maketitle

\newtheorem{prop}{Proposition}[section]
\newtheorem{claim}[prop]{Claim}
\newtheorem{theorem}[prop]{Theorem}
\newtheorem{hypo}[prop]{Hypothesis}
\newtheorem{problem}[prop]{Problem}
\newtheorem{question}[prop]{Question}
\newtheorem{remark}[prop]{Remark}
\newtheorem{lemma}[prop]{Lemma}
\newtheorem{cor}[prop]{Corollary}
\newtheorem{defn}[prop]{Definition}
\newtheorem{ex}[prop]{Example}
\newtheorem{conj}[prop]{Conjecture}
\newcommand{\ca}[1]{{\cal #1}}
\newcommand{\ignore}[1]{}
\newcommand{\f}[2]{{\frac {#1} {#2}}}
\newcommand{\embed}[1]{{#1}^\phi}
\newcommand{\stab}{{\mbox {stab}}}
\newcommand{\perm}{{\mbox {perm}}}
\newcommand{\codim}{{\mbox {codim}}}
\newcommand{\modulo}{{\mbox {mod}\ }}
\newcommand{\C}{\mathbb{C}}
\newcommand{\rel}{ \backslash}
\newcommand{\sym}{{\mbox {Sym}}}
\newcommand{\spec}{{\mbox {Spec}}}
\newcommand{\idealof}{{\mbox {ideal}}}
\newcommand{\trace}{{\mbox {trace}}}
\newcommand{\polylog}{{\mbox {polylog}}}
\newcommand{\sign}{{\mbox {sign}}}
\newcommand{\rank}{{\mbox {rank}}}
\newcommand{\poly}{{\mbox {poly}}}
\newcommand{\formula}{{\mbox {Formula}}}
\newcommand{\circuit}{{\mbox {Circuit}}}
\newcommand{\core}{{\mbox {core}}}
\newcommand{\orbit}{{\mbox {orbit}}}
\newcommand{\cycle}{{\mbox {cycle}}}
\newcommand{\ideal}{{\mbox {ideal}}}
\newcommand{\qed}{{\mbox {Q.E.D.}}}
\newcommand{\proof}{\noindent {\em Proof: }}
\newcommand{\wt}{{\mbox {wt}}}
\newcommand{\level}{{\mbox {level}}}
\newcommand{\vol}{{\mbox {vol}}}
\newcommand{\vect}{{\mbox {Vect}}}
\newcommand{\Hom}{{\mbox {Hom}}}
\newcommand{\val}{{\mbox {wt}}}
\newcommand{\adm}{{\mbox {Adm}}}
\newcommand{\eval}{{\mbox {eval}}}
\newcommand{\invlim}{{\mbox {lim}_\leftarrow}}
\newcommand{\directlim}{{\mbox {lim}_\rightarrow}}
\newcommand{\sformal}{{\cal S}^{\mbox f}}
\newcommand{\san}{{\cal S}^{\mbox an}}
\newcommand{\Oan}{{\cal O}^{\mbox an}}
\newcommand{\vformal}{{\cal V}^{\mbox f}}
\newcommand{\diffl}{\bar {\cal D}}
\newcommand{\diff}{{\cal D}}
\newcommand{\invf}{{\cal F}^{-1}}
\newcommand{\invox}{{\cal O}_X^{-1}}

\section{Abstract} \label{sintro}
In \cite{mulsoh}, henceforth  referred to as Part I, 
we suggested an approach to the $P$ vs. $NP$ and related lower bound 
problems in complexity theory  through 
geometric invariant theory. 
In particular, 
it reduces the arithmetic (characteristic zero) version of the 
$NP \not \subseteq P$ conjecture
to the problem of showing that 
a  variety  associated with the complexity class $NP$ 
cannot be embedded in a  variety associated with the complexity class
$P$. We shall call these {\em class varieties}  associated with the 
complexity classes $P$ and $NP$.

This paper  develops this approach 
further,  reducing  these lower bound problems--which 
are all  nonexistence problems--to some  existence problems: specifically
to proving  existence of  
{\em obstructions} to such embeddings among class varieties.
It gives two results towards 
explicit construction of such obstructions.

The first result is
a  generalization of the Borel-Weil theorem 
to  a class of  orbit closures, which include class varieties. 
The second result 
is a weaker form of a conjectured analogue of the second fundamental
theorem of invariant theory for the class variety associated with
the complexity class $NC$.
These  results  indicate
that the fundamental lower bound problems in complexity
theory are, in turn, intimately linked with   explicit 
construction problems in algebraic geometry and representation theory.

The results here were announced in \cite{GCToverview}. 
We are grateful to   Madhav Nori for his guidence and encouragement 
during the course of this work, and to 
C. S. Seshadri and  Burt Totaro for helpful 
discussions.

\section{Main results} \label{smainresults}
We shall now state the  results precisely. 
For the sake of completeness, we 
recall in  Section~\ref{sorbitproblem} 
the main results from part I. The rest of
this paper is self contained.
All groups in this paper are
algebraic and the base field is $\C$.

Let $G$ be a connected, reductive group,
$V$ its (finite dimensional)  linear 
representation, and $P(V)$  the corresponding 
projective space.
Let $\Delta_V[v]$ denote the projective 
closure of the $G$-orbit of $v$ in $P(V)$. It is an almost-homogeneous 
space in the terminology of \cite{akhiezer}. Let  $R_V[v]$ be its 
homogeneous coordinate ring,  $I_V[v]$ its ideal, and
$R_V[v]_d$, the degree $d$ component of $R_V[v]$.

In Part I, we reduced arithmetic (characteristic zero) 
implications  of the
lower bound problems in complexity theory,
such as $P$ vs. $NP$, and $NC$ vs. $P^{\#P}$,
to instances of
the following  (Section~\ref{sorbitproblem}):

\begin{problem} \label{pintroorbit}
 ({\em  The orbit closure problem})

Given  explicit points  $f, g \in P(V)$, 
does $f \in \Delta_V[g]$? 
Equivalently,  
is $\Delta_V[f] \subseteq \Delta_V[g]$? 
\end{problem} 

The goal is to show that this is not the case in the problems under
consideration.

The $f$'s and $g$'s here depend on the complexity classes 
in the lower bound problem under
consideration. 
In the context of the $P$ vs. $NP$ problem, the point $g$ will
correspond to a judiciously chosen $P$-complete problem, and
$f$ to a judiciously chosen $NP$-complete problem. 
We  call $\Delta_V[g]$ and $\Delta_V[f]$ the  {\em class varieties} 
associated with the complexity classes $P$ and $NP$ (this terminology
was not used in part I).
The orbit closure problem 
in this context is to show that the class variety associated with $NP$ 
cannot be embedded in a class variety associated with $P$. 
We have oversimplified the story here. There is not just 
one class variety associated with a given complexity class,
but a sequence of class 
varieties  depending on the parameters of the lower bound problem
under consideration. In the context of the $P$ vs. $NP$ problem,
the goal is to show that a class variety for $NP$ associated 
with a given  set of  parameters cannot be embedded in the 
class variety for $P$ associated with the same set of parameters. 
This would imply that $P\not = NP$ in characteristic zero.

\subsubsection*{Class variety for the complexity class $NC$}

We give an example of a class variety, associated 
with the complexity class $NC$, the class of problems with efficient
parallel algorithms. This occurs in the context of $NC$ vs. 
$P^{\#P}$ problem (Section~\ref{subspermvsdt}). Here we let $g$ be the 
determinant function, which is a complete function for this class.
Specifically, let  $Y$ be an $m\times m$ variable matrix,
which can also be thought of as a variable $l$-vector, $l=m^2$.
Let $V=\sym^m(Y)$ be the space of homogeneous forms of degree $m$ 
in the $l$ variable entries of $Y$, with the natural action of
$G=SL(Y)=SL_l(\C)$. 
Let $g=\det(Y) \in P(V)$ be the determinant form,
considered  as a point in the projective space. 
Then $\Delta_V[g]$, the orbit closure of the determinant function,
is the class variety associated with $NC$.
This is a basic example of a class variety, which the reader may
wish to keep in mind throughout this paper.

For arbitrary $f$ and $g$, Problem~\ref{pintroorbit} is hopeless.
But $f$ and $g$ 
 in the preceding lower bound problems
    can  be    chosen judiciously, like the determinant function,
 to have some special properties (cf. Section~\ref{sorbitproblem} and part I).
To state these properties, we need a few definitions.

Given   a  point $v \in P(V)$, let
 $\hat v \in V$  denote a nonzero point
on the line representing $v$; the exact choice of $\hat v$ will not matter.
Let 
$G_{v}, G_{\hat v} \subseteq G$
denote the stabilizers of $v$ and $\hat v$, respectively.
We say that  $v$
is {\em characterized}  by its  stabilizer, 
if $V^{G_{\hat v}}$, the set of points in $V$ stabilized by $G_{\hat v}$,
is equal to $\C v$,  the line in $V$ corresponding to
$v$.

Following Mumford and Kempf 
\cite{mumford,kempf}, we say  that $v$ is {\em stable} if the orbit
$G {\hat v} \subseteq V$
is closed,
and  {\em semistable}  if the closure of  this orbit
does not contain zero  \cite{mumford,kempf}. We say 
$v$ belongs to the {\em null cone}
if  all homogeneous $G$-invariants of positive degree vanish at $\hat v$.
We also define a more general notion of 
 {\em partial stability} which also applies to points in the null cone.
A stable point is also partially stable by definition.
Now suppose  $v$ is not stable. Let $S$ be any closed
$G$-invariant subset of $V$ not containing $\hat v$ and 
meeting the boundary of 
the orbit $G \hat v$.
Kempf \cite{kempf} 
associates with $v$ and $S$ a canonical parabolic subgroup 
$P=P[S,v] \subseteq G$, call its canonical destabilizing flag.
Let $L$ be its semisimple Levi subgroup.
We say that 
$v$ is {\em partially stable} with defect zero,
 or  more specifically, $(L,P)$-stable, 
if (1) the unipotent radical $U$ of $P$ is contained in
$G_v$, and (2) $v$ is stable with respect to 
the restricted action of $L$ on $V$.
A more general  notion of partial stability allowing nonzero defect 
is given later
(Definition~\ref{dpartiallystable}). 

We say that $v$ is {\em excellent} if 
\begin{enumerate} 
\item it is stable or partially stable with defect zero, and
\item it is characterized by its stabilizer.
\end{enumerate} 
If $V$ is an irreducible representation $V_\lambda(G)$ of $G$,
corresponding to a dominant weight $\lambda$,  then the point in $P(V)$
corresponding to the highest weight vector of $G$ is excellent.
This is the simplest example of an excellent point. 
In this case, the stabilizer $G_{v}$ is a parabolic subgroup $P=P_\lambda$
of $G$, and the orbit $G v\cong G/P$ is closed. Hence $\Delta_V[v]\cong G/P$.
The algebraic geometry 
of $G/P$ has been intensively studied in the literature and is
well understood by now; cf. \cite{fulton,smt} for surveys.

For the lower bound problems under consideration,
the points $f$ and $g$ can be chosen so that they are either excellent
or almost excellent; the meaning of almost excellent is stated in
Section~\ref{sorbitproblem}.
For example, the determinant function above is excellent.
In this paper, we 
shall develop an approach to the orbit closure problem specifically for such 
$f$ and $g$. 
The goal  is to understand the orbit closure problem
by systematically extending  the results for $G/P$ to the 
(almost) excellent points that arise in this approach.

A natural  approach to the orbit closure problem is the 
following. If $f$ lies in $\Delta_V[g]$, then  the 
embedding $\Delta_V[f] \hookrightarrow \Delta_V[g]$ is 
$G$-equivariant. This gives a
degree preserving 
$G$-equivariant surjection
from $R_V[g]$ to $R_V[f]$. 
Hence, if $S$ is any irreducible representation  of 
$G$, its multiplicity in $R_V[g]_d$ exceeds that 
in  $R_V[f]_d$, for all $d$. 
\begin{defn} \label{dobst1}
We  say that 
$S$ is an
 {\em obstruction} for the pair $(f,g)$ if, for some $d$,
\begin{enumerate} 
\item it occurs in (a complete $G$-decomposition of)  $R_V[f]_d$,
\item  but not in $R_V[g]_d$.
\end{enumerate} 
\end{defn} 

 Existence of such an $S$ implies that 
$f$ cannot lie in $\Delta_V[g]$. In a lower bound problem, this $S$
can  be considered to be a ``witness''
to  the computational hardness of  $f$.

If $S$ occurs in $R_V[g]_d$, then it is easy to show 
(Proposition~\ref{peval}) 
that its dual $S^*$ 
contains a $G_g$-module isomorphic to $(\C g)^d$, the $d$-th
tensor power of $\C g$. Hence:

\begin{defn} \label{dobst2}
We say that $S$ is a {\em strong obstruction} 
if, for some $d$,
\begin{enumerate} 
\item  it occurs in $R_V[f]_d$,
\item  but its dual $S^*$ does not contain 
 a $G_g$-module isomorphic to $(\C g)^d$.
\end{enumerate} 
\end{defn} 
A strong obstruction is also an obstruction.

For the $(f,g)$'s in the lower bound problems under consideration,
 strong obstructions are conjectured to
 exist in plenty (Section~\ref{sobstruction}).
But to prove their existence
it is necessary  to construct them more or less explicitly. Otherwise,
the proof technique can not cross the natural proof barrier formulated
in \cite{rudich} that any technique for proving the $P\not = NP$ conjecture
must cross. 
Explicit constructions have been used in the theory of computing 
earlier in different contexts. 
For example, explicit expanders,
needed for efficient pseudo-random generation, have been constructed 
by Margulis \cite{margulis}, and
 Lubotzky, Phillips and Sarnak
\cite{sarnak}. The essential difference from the situation here
is that proving  existence of expanders is easy, whereas proving existence 
of obstructions is itself the main problem.

Hence, we are lead to: 
\begin{problem} \label{pintroexplicit}
(Explicit Construction of  obstructions)

Given $f$ and $g$ as in Problem~\ref{pintroorbit}, explicitly construct a (strong)
obstruction for the embedding $\Delta_V[f]\hookrightarrow \Delta_V[g]$.
\end{problem}

In the orbit closure problems under consideration,
$H=G_g$ turns out to be 
a reductive subgroup of $G$. Hence, to solve Problem~\ref{pintroexplicit},
we have to solve the following problems first.

\begin{problem} \label{pintrosubgroup}
 (Subgroup restriction problem)

Let $H$ be a reductive subgroup of a connected, reductive group $G$.
Find an explicit decomposition   a given 
irreducible $G$-representation $S$ as an $H$-module.
\end{problem} 

This arises in the context of the second condition 
in Definition~\ref{dobst2}.

Problem~\ref{pintrosubgroup},
with $H$ equal to the  the stabilizer of the
determinant function considered earlier, turns out to be
equivalent to the Kronecker 
problem of finding an explicit decomposition of the
tensor product of two irreducible representations of the symmetric 
group; cf. Section~\ref{sorbitproblem}.
This is an outstanding problem in the representation theory
of the symmetric group \cite{macdonald,fulton}. 
Other 
specific instances of Problem~\ref{pintrosubgroup}  that arise in the
lower bound problems under consideration (cf. Section~\ref{sorbitproblem}) 
include the 
well known plethysm problem \cite{macdonald,fulton},
which is an outstanding 
problem in the  representation theory of $GL_n(\C)$.

\begin{problem} \label{pintrogit1} (Problem in geometric invariant theory)

Let $v\in P(V)$ be an (almost) excellent point. 

Find an  explicit 
decomposition of   $R_V[v]_d$, for a given $d$, as a $G$-module.
\end{problem} 

This is needed in the context of both conditions in Definition~\ref{dobst1}.
For this, it is desirable to solve the following problem 
first:

\begin{problem} \label{pintrogit2}  (SFT problem)

Let $v\in P(V)$ be an (almost) excellent point. 
Find an explicit set of generators for the ideal $I_V[v]$ 
of $\Delta_V[v]$ with good representation theoretic properties.
\end{problem}

Problems~\ref{pintrogit1} and \ref{pintrogit2} are
 intractable for general $v$'s. 
Hence, specialization to 
almost excellent $v$'s is necessary. Some additional 
reasonable restrictions may be necessary in these problems.

When $V=V_\lambda(G)$, 
$v$  the point corresponding to the highest weight vector 
of $V_\lambda(G)$, and  $\Delta_V[v] \cong G/P$,
the second fundamental theorem (SFT) of invariant theory for 
 $G/P$ \cite{smt}, answers 
Problem~\ref{pintrogit2}.
By the  Borel-Weil theorem for $G/P$  \cite{fulton},
$R_V[v]_d=V_{d \lambda}(G)^*$. This 
answers  Problem~\ref{pintrogit1}.

What  is desired is a  generalization of these results for $G/P$
to  the class varieties $\Delta_V[v]$, for the 
(almost) excellent $v$'s under consideration.
Before we go any further, let us point out  the main difference between 
$G/P$ and the class varieties:
\begin{enumerate} 
\item Luna and vust \cite{lunavust}
 have assigned a complexity to orbit closures,
which measures the complexity of their algebraic geometry. 
All orbit closures whose algebraic geometry has been well understood
have low Luna-Vust complexity--close to zero.
For example, the Luna-Vust complexity of $G/P$ is zero.
In contrast, the Luna-Vust complexity of a class variety can be 
polynomial in the parameters in the lower bound problem under consideration.

\item The analogue of the subgroup restriction problem 
(Problem~\ref{pintrosubgroup}), with $H$ being the parabolic 
stabilizer  $P$ of the
highest weight vector in $V_\lambda(G)$, is trivial.
In contrast, the instances of Problem~\ref{pintrosubgroup}
 in the context of the
class varieties include the nontrivial Kronecker and plethysm problems.
\end{enumerate} 

This indicates that the  algebraic  geometry of class varieties  is
substantially more complex  than that of $G/P$. 
For this reason, we cannot expect a full solution to 
Problems~\ref{pintrogit1} and \ref{pintrogit2}  until the outstanding 
Problem~\ref{pintrosubgroup} in representation theory  is resolved.
Rather, our goal is to connect  Problems~\ref{pintrogit1} and \ref{pintrogit2}
  with the ``easier'' Problem~\ref{pintrosubgroup} for the almost 
excellent $v$'s under consideration.
We prove two results in this direction.

Let us begin by considering a weaker
form of Problem~\ref{pintrogit1}; i.e., we only ask 
which $G$-modules can occur in $R_V[v]$, without worrying about
$R_V[v]_d$ for a specific $d$. This is addressed  by the following result.

We call a $G$-module $V_\lambda(G)$ $G_{\hat v}$-admissible if 
it contains a $G_{\hat v}$-invariant (cf. Definition~\ref{dadmissible}).

\begin{theorem} \label{tlieovernew}
(Borel-Weil for orbit closures of partially stable points)

Let $V$ be a (finite dimensional) 
linear representation of a connected, reductive $G$.

\noindent (a) If $v \in P(V)$ is stable, 
an irreducible  $G$-module $V_\lambda(G)$ with weight $\lambda$ can occur in
$R_V[v]$ iff 
$V_\lambda(G)$ is $G_{\hat v}$-admissible.

\noindent (b) 
Suppose $v$ is partially stable with defect zero,
specifically  $(L,P)$-stable, as defined above.
Let $S_V[v]$ be the homogeneous 
coordinate ring  of the  projective closure in $P(V)$  of the $L$ orbit of $v$.
Then the $G$-module structure of $R_V[v]$ is completely determined 
by  the $L$-module structure of $S_V[v]$.
A weaker statement holds a partially stable point of nonzero defect
as defined in Section~\ref{spartiallystable}. 
\end{theorem} 

A precise statement of (b) is given in 
Section~\ref{subspstable}. 
We  actually prove a stronger result 
(Theorem~\ref{tlieoverrefined}) that 
specializes to the Borel-Weil theorem \cite{knapp}
when $v$ corresponds to the highest weight vector of an irreducible 
representation $V=V_\lambda(G)$ of a semisimple $G$.

When the defect is nonzero, Theorem\ref{tlieovernew} (b) 
does not  tell precisely which irreducible $G$-modules
occur in $R_V[v]$  if we only knew which irreducible $L$-modules
occur in $S_V[v]$ as a whole. But it gives a good information 
on this and also on which irreducible $G$-modules 
occur in $R_V[v]_d$, for a given $d$,
provided we   know  precisely which irreducible 
$L$-modules occur in every degree $d$-component $S_V[v]_d$;
this is Problem~\ref{pintrogit1}  for a stable $v$, with $L$ playing 
the role of $G$.

Now we turn to the actual 
Problem~\ref{pintrogit1}.
For this, we have to understand 
Problem~\ref{pintrogit2} first.
We turn to this problem next. 

Let $v$ be an excellent point.
We associate with it 
a representation-theoretic 
data $\Pi_v=\cup_d \Pi_v(d)$ (cf. Definitions~\ref{defnnonadmissibledata_new}
and \ref{defnnonadmissibledata_new2}). If 
$v$ is stable, $\Pi_v(d)$ 
is just the set of all irreducible $G$-submodules
of $\C[V]$ whose duals do not contain a $G_v$-submodule isomorphic 
$(\C v)^d$.

Then the ideal $I_V[v]$ contains all modules in $\Pi_v$ 
 (Proposition~\ref{peval}).
Let $X(\Pi_v)$ be the variety (scheme) defined by the ideal 
generated by the modules in $\Pi_v$.
It follows that  $\Delta_V[v] \subseteq X(\Pi_v)$. 

Now we ask: 
\begin{question} \label{qintroconj}
Suppose $v$ is excellent.
Is  $X(\Pi_v)=\Delta_V[v]$ as a variety, or more strongly,
as a scheme?
\end{question} 

The scheme theoretic equality  means that the ideal $I_V[v]$ of 
$\Delta_V[v]$ is generated by the modules in $\Pi_v$.

If $v$ is stable, then $G_v$ is reductive \cite{borel,matsushima}. Hence,
the $G$-modules contained in  $\Pi_v$ are precisely determined 
once we know answer to Problem~\ref{pintrosubgroup}, with $H=G_v$. 
This turns out to be so even for the  partially stable $v$'s that arise
in the lower bound problems, by letting  $H$ be the reductive part of $G_v$. 
Hence, if the answer to Question~\ref{qintroconj} is yes, 
the algebraic geometry of $\Delta_V[v]$ is completely determined
by the representation theory of the pair $(G_v,G)$, and hence, 
Problems~\ref{pintrogit1} and \ref{pintrogit2}  are intimately related to
Problem~\ref{pintrosubgroup}.
Clearly, this can happen only for very special $v$'s. The
answer need not be yes even for a general excellent $v$.

When $v$ corresponds to the highest weight vector of $V_\lambda(G)$,
so that $\Delta_V[v]=G/P$,
answer to Question~\ref{qintroconj} is yes.
This follows from the second fundamental
theorem (SFT) for $G/P$ \cite{smt}  (cf. Section~\ref{sgmodp}).  

We conjecture that this is also the case  for the class variety
associated with the complexity class $NC$ described above.

\begin{conj} \label{conjsftdet} (Second fundamental theorem (SFT) for the 
orbit closure of the determinant) 

Let $\Delta_V[v]$ be the class variety associated with the
complexity class $NC$--the orbit closure of the determinant  function.

Then $X(\Pi_v)=\Delta_V[v]$ as a variety, and more strongly,
as a scheme. 
\end{conj} 

This is expected because of  the very special nature of the 
 determinant function. 
We have already remarked that it is excellent.
Furthermore, its stabilizer has an additional conjectural property
called $G$-separability (Definition~\ref{dgseparable}).
For analogous conjectures for  other almost excellent class varieties,
 one has to address complications caused by 
 almost excellence instead of full excellence. This is possible,
and will be done elsewhere.

The following general result   implies  a weaker form 
of   Conjecture~\ref{conjsftdet} when $v$ is the determinant function.

\begin{theorem} \label{tnonadmissibilitynew} 
(Second Fundamental Theorem (SFT) for the orbit of an excellent 
 point)

Suppose $V$ is a linear  representation of a connected, reductive group 
$G$, and  $v\in P(V)$ an excellent point.

\noindent (a) Suppose $v$ is stable. 
Furthermore, assume that the  stabilizer $G_{\hat v}$  is $G$-separable
 (cf. Definition~\ref{dgseparable}).
Then the orbit $G v \subseteq P(V)$
is determined by the representation-theoretic 
data  $\Pi_v$ within some $G$-invariant neighbourhood $U$: i.e.,
\[ G v =\Delta_V[v] \cap U = X(\Pi_v) \cap U,\]
as schemes. 

\noindent (b) A generalized result also holds for the $G$-orbit of a
partially stable, excellent  point with defect zero.
\end{theorem}

This follows from a stronger result proved in Section~\ref{subsidealstable}
(stable case)
and Section~\ref{subsidealpstable} (partially stable case). 

When $v$ corresponds to the highest weight vector in $V_\lambda(G)$,
Theorem~\ref{tnonadmissibilitynew} (b), after some strengthening
(cf. Section~\ref{sgmodp}),  becomes the 
second fundamental theorem for $G/P$ \cite{smt}--hence the terminology.

The rest of this paper is organized as follows. 
In section~\ref{sorbitproblem},
we describe how  the orbit closure problem arises in
complexity theory, and summarize 
the relevant results from part I.
In Section~\ref{sadmissible} we prove some basic propositions 
based on the notion of admissibility. 
The stable case of Theorem~\ref{tlieovernew}
 is proved in Section~\ref{subsstable}.
The stable case of Theorem~\ref{tnonadmissibilitynew}
 is proved in Section~\ref{subsidealstable}.
The stable cases illustrate the main ideas in this paper.
The notion of partial stability is introduced in 
Section~\ref{spartiallystable}.
The partially stable case of Theorem~\ref{tlieovernew}
 is proved in Section~\ref{subspstable}.
Its specialization in the context of complexity theory is given
in Section~\ref{scomplexity}. 
The partially stable case of Theorem~\ref{tnonadmissibilitynew} 
is proved in Section~\ref{subsidealpstable}.
Conjectural $G$-separability of the stabilizer of the determinant
is proved in Section~\ref{sseparable} for a special case.

\subsubsection*{Notation} 
We let $G$  denote  a connected reductive group.
An irreducible  $G$-representation 
 with  highest weight $\lambda$ will be
 denoted by $V_\lambda(G)$. We say that $V_\lambda(G)$ occurs in 
a  $G$-module $M$, or that $M$ contains $V_\lambda(G)$,
 if a complete decomposition of $M$ into $G$-irreducibles
contains a copy of $V_\lambda(G)$. We denote the dual of  $M$ by $M^*$.
We  always denote a Levi decomposition of a parabolic subgroup
$P \subseteq G$ in the form
 $P=TLU=KU$,  where $T$ is a torus,
$L$ is a semisimple Levi subgroup, $K=TL$ is a reductive  Levi subgroup,
and $U$ is the unipotent radical.
The root system of  $K$ is a subsystem of that of $G$. Hence
a dominant weight of $G$ can be assumed  to be a dominant weight of $K$
by restriction.

\section{The orbit closure problem} \label{sorbitproblem}
In this section we describe the orbit closure problem that arises in
complexity theory, and the related results; cf. Part I for  details and
proofs.

Let $Y=[y_0,\cdots,y_{l-1}]$ denote a variable  $l$-vector.
For $k<l$, let $X=[y_1,\cdots,y_k]$, and $\bar X=[y_0,\cdots,y_k]$ be its 
 subvectors of size $k$ and $k+1$.
Let $V=\sym^m(Y)=\sym^m((\C^{l})^*)$ be 
the  space of homogeneous forms of degree $m$ in the $l$ variable-entries of
$Y$,
 with the natural action of 
$G=SL(Y)=SL_l(\C )$, and $\hat G=GL(Y)=GL_l(\C)$.

Let  $W=\sym^{n}(X)$, $n < m$, be the representation of $GL(X)=GL_k(\C)$.
We have a natural embedding $\phi: W \rightarrow V$, which maps 
 any $w\in W$ to $y^{m-n}w$, where $y=y_0$ is used as the
homogenizing variable.  The image $\phi(W)$  is contained in 
$\bar W=\sym^m(\bar X)$, a representation of $GL(\bar X)=GL_{k+1}(\C)$.

\begin{defn}\label{defn}  \label{dpstablebasic}
We say that $f=\phi(h)$ is partially stable with respect to the action
of $G$ if 
$h \in P(W)$ is stable with respect to the action of $SL_k(\C)$.
\end{defn} 
These are the only kinds  of partially stable points that arise
in the context of complexity theory. If the reader wishes,
he may confine himself to only these kinds.
When we introduce   a more general definition of partial stability 
(Section~\ref{spartiallystable}),
it will turn out that $f$ is partially stable with
defect one. In contrast, the $(L,P)$-stable points in the introduction
will turn out to be partially stable points with defect zero.
Note that $f$ in Definition~\ref{dpstablebasic}  belongs to the 
null cone of the $G$-action--this  follows easily from the Hilbert-Mumford
criterion \cite{mumford}.

The orbit closure problems (Problem~\ref{pintroorbit}) 
that arise in complexity theory (cf. Part I) have 
the following form:
\begin{problem} \label{porbitclosure}
Given  fixed forms $g \in P(V)$,
and $h \in P(W) \stackrel {\phi} {\hookrightarrow} P(V)$,
does $f=\phi(h)$ belong to  $\Delta_V[g]$? That is,
is $\Delta_V[f] \subseteq \Delta_V[g]$?
\end{problem}
The goal is to show that the specific $f$ does not belong to $\Delta_V[g]$.
The specific $f$ and $g$ depend on the lower bound problem under 
consideration, and will be either excellent (cf. Section~\ref{sintro}), or
almost excellent--the latter  means: (1) the defect of partial stability may
not be zero, but will be small, and (2) the point 
may not  be fully characterized by the stabilizer, but almost (as 
explained in part I).

The following are two instances of the orbit closure problem that arise in
complexity theory. 

\subsection{Arithmetic  version of the 
$NC$ vs.  $P^{\#P}$  conjecture} \label{subspermvsdt}
In concrete terms, this says that 
the permanent of an $n \times n$ matrix cannot be computed by 
an integral circuit of depth $\log^c n$, for any constant $c>0$
\cite{valiant}. 

The class varieties in this context are as follows.
Let $Y$ be an $m\times m$ variable matrix,
which can also be thought of as a variable $l$-vector, $l=m^2$. Let
$X$ be its, say,  principal bottom-right 
$n\times n$ submatrix, $n<m$, which can be thought of as a 
variable $k$-vector, $k=n^2$. We use  any entry $y$  of $Y$ not in 
$X$ as the homogenizing variable for embedding $W = \sym^n(X)$ in 
$V=\sym^m(Y)$.
Let $g=\det(Y) \in P(V)$ be the determinant form (which will also be
considered  as a point in
the projective space), and $f=\phi(h)$, where 
$h=\perm(X) \in   P(W)$. 
Then $\Delta_V[g]$ is the class variety associated with
$NC$ and $\Delta_V[f]$ the class variety associated with 
$P^{\#P}$. These depend on the lower bound parameters $n$ and $m$.
If we wish to make these implicit, we should write $\Delta_V[f,n,m]$ 
and $\Delta_V[g,m]$ instead of $\Delta_V[f]$ and $\Delta_V[g]$. 

It is conjectured in part I that, if $m=2^{O(\polylog n)}$ and
 $n \rightarrow \infty$, then 
$f \not \in \Delta_V[g]$; i.e., 
the class variety $\Delta_V[f,n,m]$ cannot be embedded in the class 
variety
$\Delta_V[g,m]$. This implies the arithmetic form of the
$NC \not = P^{\#P}$ conjecture.

The following result provides the connection with geometric invariant theory.

\begin{theorem} (cf. Part I) \label{tpart11}
The point  $h=\perm(X) \in P(W)$ 
is stable with respect to the action of $SL(X)=SL_k(\C )$ 
on $P(W)$ (thinking of $X$ as a $k$-vector). Hence the point
$f=\phi(h) \in P(V)$ is partially stable (definition~\ref{dpstablebasic})
with respect to the
action of $G=SL(Y)=SL_l(\C)$,  as well as $\hat G=GL(Y)=GL_l(\C)$.

Similarly, $g=\det(Y) \in P(V)$ 
is stable with respect to the action of $G$
on $P(V)$, thinking of $Y$ as an $l$-vector on which $SL_l(\C)$ acts in
the usual way. 
\end{theorem}

Moreover, both  $\perm(X) \in P(W)$ and $\det(Y) \in P(V)$
are characterized by their stabilizers. Hence, both $h$ and
$g$ are excellent. But, in contrast, $f=\phi(h)$ is only 
almost excellent--because  its defect of partial stability is one.

The stabilizer of $\det(Y)$ in $G=SL_{m^2}(\C)$ consists of 
linear transformations  of the form
$Y \rightarrow A Y^{*} B^{-1}$,  thinking of $Y$ as an $m\times m$ matrix,
where $Y^*$ is either $Y$ or $Y^T$,
 $A,B \in GL_{m}(\C)$.
  The stabilizer of $\perm(X)$ in $SL_{n^2}(\C)$
is generated \cite{minc} by linear transformations of the 
form 
$X \rightarrow \lambda X \mu^{-1}$, 
thinking of $X$ as an $n \times n$ matrix,
where $\lambda$ and $\mu$ are either diagonal or permutation
matrices.

Let $H\subseteq G=SL_{m^2}(\C)$ be the stabilizer of $\det(Y)$. 
Since  $SL_m(\C)\times SL_m(\C)$ is a subgroup of $H$, the 
subgroup restriction problem (Problem~\ref{pintrosubgroup})
in this context becomes:

\begin{problem} {\em (Kronecker problem)}

Given a partition $\lambda$ of height $m^2$, find an explicit 
decomposition of $V_\lambda(G)$ as an $SL_m(\C) \times SL_m(\C)$-module: 
\[ 
V_\lambda(G)=\oplus_{\alpha,\beta} k_{\alpha,\beta}^\lambda 
V_\alpha(SL_m(\C)) \otimes 
V_\beta(SL_m(\C)),
\] 
where $\alpha,\beta$ range over partitions of height at most $m$.
\end{problem}
The coefficients $k_{\alpha,\beta}^\lambda$'s here are the same as the
Kronecker coefficients that arise in the internal product of Schur
functions.
The problem of decomposing the tensor product of two irreducible 
representations of the symmetric group $S_m$ can be reduced to this
problem
\cite{fulton}. This is one of the outstanding problems in the representation
theory of symmetric groups.

\subsection{Arithmetic  (nonuniform) version of the 
$P \not = NP$ conjecture} \label{subsecpvsnp}
This is a version of the usual  $P \not = NP$ conjecture (the 
nonuniform version), which does not involve problems of positive
characteristic, and hence, is addressed first.

Now $h,g$ in the orbit closure problem (Problem~\ref{porbitclosure}) 
correspond to some integral
functions that
are $NP$-complete and $P$-complete, respectively. These functions
have to be chosen judiciously, because
most functions that arise in complexity theory,
e.g. the one associated with the travelling salesman problem, do not
have a nice stabilizer,
as  required in our approach. For a detailed definition of 
$h$ and $g$, see part I. 
We shall call $\Delta_V[f]$, $f=\phi(h)$, and $\Delta[g]$ for the
specific $h$ and $g$ here the {\em class varieties} associated with
the complexity classes $NP$ and $P$.   The conjecture that
$NP \not \subseteq P$ in characteristic zero is then
reduced to the problem 
of showing that the class variety $\Delta_V[f]$ associated with $NP$ cannot be
embedded in the class variety $\Delta_V[g]$ associated with $P$,
for the parameters of the lower bound problem under consideration.

The following is an analogue of Theorem~\ref{tpart11} in this context.

\begin{theorem} \label{tpart12}
The point $h\in P(W)$, for a suitable $W$, 
which  corresponds to an $NP$-complete function as in \cite{mulsoh}, 
is stable with respect to the action of 
$SL(W)$ on $P(W)$. Hence, the point $f=\phi(h)$ is partially stable.
\end{theorem}

The $h$ here  is  not completely characterized 
by its stabilizer, but almost so; cf. part I. Hence it is almost excellent.
The subgroup restriction problem Problem~\ref{pintrosubgroup}  that arises
for the stabilizer of $h$ 
is essentially  the well known plethysm problem \cite{fulton} in the theory of
symmetric functions.

\section{Why should obstructions exist?} \label{sobstruction}
Before we go any further, we have to argue why obstructions
should exist for the pairs $(f,g)$ that arise in the lower bound 
problems under consideration.

Let us begin with an observation  that for an orbit closure problem 
that arises in complexity theory,
an obstruction for the pair $(f,g)$ cannot exist if $l$ is 
sufficiently larger than $k$.
For example, let  $(f,g)=(\phi(h),g))$, where $h=\perm(X)$ and 
$g=\det(Y)$, as in Section~\ref{subspermvsdt}.
 Then there cannot be any obstruction for
$m > n!$, or for that matter, $m > 2^{c n}$ for a large enough constant $c$.
 This is because 
$\perm(X)$ has a formula of size $2^{ c n}$ for a large enough $c>0$
 \cite{minc} (the
 usual formula is of size $n!$) and
 hence $f \in \Delta_V[g]$, for $m > 2^{ c n}$ (cf. Part I).

At the other extreme, when $l=k$, so that $f$ is a stable point of $V$,
it follows from  the  \'etale slice theorem \cite{mumford,luna} 
that, if $f \in \Delta_V[g]$, then some conjugate of the stabilizer of 
$f$ must be contained in the stabilizer of $g$ (cf. Part I). 
This will not happen for our judiciously chosen  $f$ and $g$.
For example, when $f$ and $g$ are the permanent and the determinant and
$m=n$--in fact, in this case, there are infinitely many obstructions
to this containment (cf. part I).

The goal is to understand  the transition between these two extremes.

First,  let us consider the arithmetic implication of the 
$P^{\#P} \not = NC$ conjecture. Let $g=\det(Y)$, $f=\phi(h)$, and  $h=\perm(X)$
as in Section~\ref{subspermvsdt}.

\begin{prop} \label{pconjexobs}
Suppose $h=\perm(X)$ cannot be approximated infinitesimally  closely by 
 a circuit of
depth $O(\log^{c} n)$, where $c>0$ is a constant, and $n \rightarrow \infty$.
 Suppose $X(\Pi_g)=\Delta_V[g]$ as varieties
(cf. Conjecture~\ref{conjsftdet}). 
Then  there exists 
a strong  obstruction for the pair $(f,g)$, for $m \le 2^{\log^{c/2}n}$
\end{prop} 
\proof It is proved in Part I that the hypothesis implies that 
$f \not \in \Delta_V[g]$ if $m \le 2^{\log^{c/2}n}$. Assuming 
$X(\Pi_g)=\Delta_V[g]$, this means $f \not \in X(\Pi_g)$. Hence
 there exists a $G$-module 
 $S \in \Pi_g$ which does not vanish on $f$, and hence on its orbit. So
 $S$ occurs in $R_v[f]$. By the definition of $\Pi_g$,
the dual $S^*$ does not contain 
 a $G_g$-module isomorphic to $(\C g)^d$.
Hence $S$ is  a strong  obstruction for the pair $(f,g)$. \qed

Since $\perm(X)$ is $\#P$-complete \cite{valiant}, it is not expected to have
infinitesimally close  approximations by circuits of $O(log^c(n))$ depth,
for any constant $c>0$.
Hence, Proposition~\ref{pconjexobs} leads to:

\begin{conj} 
There exist (infinitely many) strong obstructions  for $(f,g)=(\phi(h),g)$,
$g=\det(Y)$, $h=\perm(X)$, if 
$m=2^{\log^c n}$, $c$ a constant, and $n \rightarrow \infty$. 
\end{conj}

In turn, this conjecture implies 
 $f \not \in \Delta_V[g]$, and hence,
 the arithmetic implication of the $P^{\#P} \not = NC$ conjecture
(Section~\ref{subspermvsdt}).

In the same vein, we also make:

\begin{conj} \label{conjobpvsnp}
There exist (infinitely many) obstructions  for $(f,g)=(\phi(h),g)$,
that occur in the context of the $P$ vs. $NP$ problem, if 
$m=\poly(n)$,  and $n \rightarrow \infty$, where $n$ denotes the
input size parameter, and $m$ denotes the circuit size parameter
in the nonuniform version of the $P$ vs. $NP$ problem.
\end{conj}
This would imply $f \not \in \Delta_V[g]$, and hence, the
arithmetic implication of the $P \not =NP$ conjecture
in Section~\ref{subsecpvsnp}. This conjecture 
is motivated by similar considerations as in Proposition~\ref{pconjexobs}.
The $g$ that occurs in the context of the $P$ vs. $NP$ problem is
not fully characterized by its stabilizer. But it is still determined by 
its stabilizer to a large extent. Hence, similar considerations apply.

\section{Admissibility} \label{sadmissible}
In this section, we introduce a basic notion of
 admissibility  and study how it influences which $G$-modules 
may appear in the homogeneous coordinate ring $R_V[v]$ of the 
projective-orbit closure of a point $v \in V$.
The basic propositions proved here will be useful in the proofs of
 the main results.

\begin{defn} \label{dadmissible}
Given a reductive subgroup  $H \subseteq G$ and an $H$-module $W$,
we say that a $G$-module $M$ is $(H,W)$-admissible, if some irreducible
$H$-submodule of $M$  occurs in $W$.

We say that $M$ is $H$-admissible if it is $(H,1_H)$-admissible, where
$1_H$ is the trivial $H$-module; i.e., if it contains a (nonzero) 
 $H$-invariant.

For general $H$, not necessarily reductive,
 we say that $M$ is $H$-admissible if 
$M^*$ contains an $H$-invariant.
\end{defn} 

If $H$ is reductive,  $M$ 
contains an $H$-invariant, iff $M^*$ does--this follows from Weyl's result on
complete reducibility of a reductive group representation--and hence,
the second and third statement are then equivalent.

Given a $G$-module $S$, and a subgroup $H \subseteq G$, not necessarily 
reductive, we shall say that $S$ has an $H$-coinvariant if 
$S$ is $H$-admissible, i.e., 
the dual module $S^*$ has an $H$-invariant (cf. Definition~\ref{dadmissible}).

Let  $h \in P(V)$ be any point, not necessarily stable.
Let  $\C h$ be the corresponding line in $V$. It 
 is  one-dimensional, i.e., a character, as a $G_h$-module , and 
trivial as a 
 $G_{\hat h}$-module. 
 Let $\check \Delta[h] = \check \Delta_V[h] 
\subseteq V$
denote the  affine cone of the projective-orbit-closure $\Delta[h]$. 
Its coordinate ring  $\C [\check\Delta[h]]$ coincides with the 
homogeneous coordinate ring  $R[h]=R_V[h]$ of $\Delta[h]$. 
Since the $G$-action is degree preserving, each 
homogeneous component  $R[h]_d$ is a finite dimensional 
$G$-module.

\begin{prop} \label{peval} 
Let $V$ be a linear representation of a reductive group $G$.
Let $h\in P(V)$ be any point, not necessarily stable, with stabilizer 
$G_h \subseteq G$. 
Let $S$ be any 
 irreducible 
 $G$-module  occurring in $R[h]$--that is,
 in  $R[h]_d$ for some $d$. Then the dual module $S^*$ must contain 
 a $G_h$-submodule isomorphic to $(\C h)^d$, and hence both 
 $S$ and $S^*$ are   $G_{\hat h}$-admissible.
 In particular, a $G$-module $S \subseteq \C[V]_d$
not satisfying this constraint  belongs to the ideal $I_V[h]$.

Similarly, given an algebraic subgroup $H \subseteq G$, and 
an $H$-module $M$, let $B = G\times_H M$ be the induced bundle 
\cite{mumford} with base space $G/H$ and fibre $M$.
Let $N$ be any  irreducible $G$-submodule of
 $\Gamma(G/H, B)$, the space of global sections of $B$. Then 
the $G$-module 
$\Hom(N,M)$ must contain a nonzero  $H$-invariant. 
\end{prop}
\proof 
Not all functions in $S$ can vanish at $\hat h$: Otherwise,
they will vanish identically on the $G$-orbit of $\hat h$ in $V$,
 and so also on its cone,
since the functions in $S$ are homogeneous. But the cone of the 
affine $G$-orbit
 of  $h$ is dense in $\check \Delta[h]$. Hence, it would follow that 
the  functions in $S$ vanish on $\check \Delta[h]$ identically; a 
contradiction.

Consider the $G_h$-equivariant  map   $\phi:  S \rightarrow 
((\C h)^*)^d = (\C h^d)^*$
  that maps every function in $S$ to its
restriction on the line $\C h$. 
 It follows that this  evaluation map is nonzero. Hence
the dual map $\phi$ injects the $G_h$-module $\C h^d$ into $S^*$.
\ignore{
Similarly it follows  that $S^*$  contains a $G_{\hat h}$-invariant, by
considering  the evaluation map $\phi_{\hat h}: S \rightarrow \C$,
which maps every function in $S$ to its value at $\hat h$.}

The argument  extends to the vector bundle $B$ by considering 
instead the evaluation map $\phi: N \rightarrow M$ at the base point $e
\in G/H$,
 which 
must be  nonzero and $H$-equivariant; i.e., $\phi \in \Hom(N,M)^H$.
\qed

\section{Admissibility and  stability} \label{subsstable}
In this section we shall prove the first statement of 
Theorem~\ref{tlieovernew}  concerning  stable points.

\begin{prop} \label{pstableorb}
Let $h \in P(V)$ be a point such that
 the stabilizer $H = G_{\hat h}$
 of $\hat h \in V$ is reductive. Then every irreducible
 $G$-module  occurring in 
 $R[h]$ must be $H$-admissible; i.e., must contain a nonzero
$H$-invariant.

If $G_{\hat h}$ is not reductive, this still holds if $H$ is any reductive 
subgroup of $G_{\hat h}$.
\end{prop} 
\proof 
If $H$ is reductive,
then Weyl's theorem on complete decomposibility of $H$-modules
into irreducibles implies that the existence of an $H$-invariant is equivalent
to existence of an
$H$-coinvariant. Hence this follows from Proposition~\ref{peval}. \qed

\noindent
Conversely,
\begin{prop} \label{pstableorbit2}
Suppose $h \in P(V)$ is stable. Then every $H$-admissible, irreducible 
$G$-module occurs in
 $R[h]$.
\end{prop}
\proof 
Since $h$ is stable, the stabilizer $H=G_{\hat h}$
 is reductive \cite{harish,matsushima},
 and the
orbit $G \hat h \subseteq V$  is affine and isomorphic to 
$G/H$ \cite{mumford}. Moreover, an explicit $G$-module decomposition 
of the coordinate ring 
$\C [G \hat h] = \C [G/H]$ can be computed as follows. First,
we recall (cf. Page 48, \cite{springer}) the algebraic version of the

\noindent {\bf Peter-Weyl Theorem:}
\begin{equation} \label{eqpeterwe} 
 \C [G] = \oplus_S S \otimes S^*,
\end{equation} 

where $S$ ranges over all irreducible
 $G$-modules,  and $S^*$ is the dual module.
\ignore{(This is equivalent to 
the usual form of the Perter-Weyl theorem applied to the Hilbert space 
$L_2(D)$ of $L_2$-functions on a maximal compact subgroup $D \subseteq G$; 
this follows from the fact that 
 $\C [G]$ coincides with the space of so-called 
finite $D$-vectors 
within $L_2(D)$ \cite{carter}).}
From this it follows that
\begin{equation} \label{epeterweyl}
 \C [G/H] = \oplus_S S \otimes (S^*)^H,
\end{equation} 
where $(S^*)^H$ denotes the subspace  of $H$-invariants in $S^*$.
Since $h$ is stable, the affine orbit $G \hat h$ is  closed in $V$.
So it is a closed $G$-subvariety of 
the cone $\check \Delta[h] \subseteq V$,
 which is also a $G$-variety. It follows that
there is a $G$-equivariant surjection from 
$R [h]$ to $\C [G \hat h]= \C [G/H]$.
 Both $R [h]$ and $\C [G/H]$ have direct sum
decompositions into finite dimensional $G$-modules. 
 It  follows that 
every irreducible $G$-module that occurs in $\C [G/H]$ must occur in 
 $R [h]$. But by the Peter-Weyl theorem, i.e.,
 eq(\ref{epeterweyl}),
the irreducible $G$-modules that appear within $\C [G/H]$ are precisely the
$H$-admissible ones. \qed 

\noindent {\bf Proof of Theorem~\ref{tlieovernew} (a):} 
Since $v \in P(V)$ is stable, $G_{\hat v}$ is reductive
 \cite{harish,matsushima}. Hence this follows from
 Propositions~\ref{pstableorb} and \ref{pstableorbit2}.
 \qed 

\section{SFT for the orbit of a stable, excellent
 point} \label{subsidealstable}
In this section we 
shall now prove  Theorem~\ref{tnonadmissibilitynew} for stable points.
To give its precise statement, we need a few definitions.

We associate with a stable point $v$ 
representation-theoretic data $\Pi_v$ and $\Sigma_v\subseteq \Pi_v$ as follows.

\begin{defn} \label{defnnonadmissibledata_new}
Suppose  $v \in P(V)$ is stable.

Let $\Sigma_v$ be the set of all 
non-$G_{\hat v}$-admissible $G$-submodules of $\C[V]$-- here $G_{\hat v}$ is 
necessarily reductive \cite{harish,matsushima}.

Let  $\Pi_v=\cup_d \Pi_v(d)$, where
 $\Pi_v(d)$ is  the set of all irreducible $G$-submodules
of $\C[V]$ whose duals do not contain a $G_v$-submodule isomorphic 
$(\C v)^d$--the $d$th tensor power of $\C v$.
\end{defn} 

Clearly $\Sigma_v \subseteq \Pi_v$. Basis elements (suitably chosen) 
of the $G$-submodules of
$\Sigma_v$ will be called {\em nonadmissible basis elements}.

\begin{prop} \label{pcnonadideal_new1}
If $v$ is stable, the  $G$-modules in
the representation-theoretic data $\Pi_v$, and hence $\Sigma_v$,
associated with $v$ are
contained in $I_V[v]$.
\end{prop} 
This follows from Proposition~\ref{peval}.

\begin{defn} \label{dgseparable}
Given a reductive $H \subseteq G$,
we say that a nontrivial, irreducible  $H$-module $L$, which occurs in
some $G$-module,  is $G$-separable 
(from the trivial $H$-module) if 
there exists an irreducible non-$H$-admissible 
  $G$-module $M$ that contains $L$; we say it is strongly $G$-separable 
if there exist infinitely many such $G$-modules.
We shall say that a subgroup $H \subseteq G$ is $G$-separable (strongly
$G$-separable), if every
nontrivial irreducible $H$-module, which occurs in some $G$-module,
  is $G$-separable (resp. strongly $G$-separable).
\end{defn}

For example, $SL_k(\C) \subseteq SL_n(\C)$, $k > n/2$, 
and  a semisimple
$H \subseteq H \times H$ (diagonal embedding) are separable
(Proposition~\ref{pbasicsep}). We conjecture that
$SL_n(\C) \times SL_n(\C)  \subseteq SL(\C^n \otimes \C^n) = SL_{n^2}(\C)$
 is separable, and prove this
 for $n=2$ (Proposition~\ref{psl2}).
We also conjecture that 
 the stabilizers of the permanent, the determinant
and other functions that arise in our lower bound applications 
 are $G$-separable; the stabilizer of the determinant is very similar to
the subgroup 
$SL_n(\C) \times SL_n(\C)  \subseteq SL_{n^2}(\C)$
above (cf. Section~\ref{subspermvsdt}).

A precise statement of Theorem~\ref{tnonadmissibilitynew} now reads
as follows:

\begin{theorem} \label{tnonadmissibility_new2}
Suppose $V$ is a linear  representation of a connected, reductive group 
$G$. Let $v \in P(V)$ be a stable point 
 such that  stabilizer $G_{\hat v}$  is $G$-separable 
(cf. Definition~\ref{dgseparable}) and characterizes $v$.

Then
there exists a homogeneous $G$-invariant $\beta \in \C[V]$ not vanishing at $v$
such that
the ideal of $G v$ as a closed subvariety of the open neighbourhood 
$U=V_\beta = V \setminus 
\{\beta=0\}$ is generated by the nonadmissible  basis elements 
in $\Sigma_v$--in fact,
it is generated by the basis of less than
$\codim(G v, P(V))$) irreducible, non-$G_{\hat v}$-admissible $G$-submodules 
of $\C[V]$.
\end{theorem}

\noindent {\bf Remark:} Since $\Sigma_v \subseteq \Pi_v$, this
statement is slightly stronger than Theorem~\ref{tnonadmissibilitynew}.

Theorem~\ref{tnonadmissibility_new2},
in turn,    follows from the following stronger result.

Let $X$ be a nonsingular, affine
 $G$-variety, $G$ a connected reductive group.
Given a point $x \in X$, we shall denote by $[x] \subseteq X$
 the subvariety consisting 
of all points in $X$ whose stabilizers contain $H=G_x$, the stabilizer of $x$.
 Assume that 
$x$ is a nonsingular point of $G \cdot [x]$; when the orbit $G x \subseteq X$
is closed, this  is automatically so, because of the \'etale slice theorem
\cite{luna} (cf. the proof of Lemma~\ref{lfirststep} below.)
We shall denote by $N_x$ (resp. $N_{[x]}$)  the $H$-module that is 
an $H$-complement 
of the tangent space of $G \cdot x$ (resp. $G \cdot [x]$) at $x$  in the total 
tangent space to $X$ at $x$; it  can be thought of as the
``normal'' space to $G \cdot x$.  (resp. $G \cdot [x]$) at $x$.
$N_{[x]}$ is the $H$-submodule of $N_x$ consisting of  all nontrivial 
$H$-submodules of $N_{x}$.

Given a $G$-invariant $\beta \in \C[X]$,
 we shall denote by $X(\beta)$ the $G$-variety 
obtained from $X$ by removing the divisor $\{\beta=0\}$.

We shall denote the codimension of a subvariety $Y \subseteq X$ by
$\mbox{codim}(Y,X)$. We say that an open subset $U \subseteq X$ is 
saturated if its of the form $\psi^{-1}(U')$, where $\psi$ is the projection
from $X$ to $X/G$ and $U'$ is an open subset of $X/G$.

 Theorem~\ref{tnonadmissibility_new2} for stable points in $P(V)$
 follows from the following result by letting $X=V$ and $x=\hat v$.
When  $v \in P(V)$ is characterized by the stabilizer $G_{\hat v}$,
 $[x]= \C v$
 Passage from $V$ to $P(V)$ is 
possible because the nonadmissible basis elements are homogeneous.

\begin{theorem} \label{tnonadmrel}
Assume that $G$ is a connected, reductive group, and $X$ an 
affine, nonsingular, irreducible   $G$-variety $X$.
Let $x \in X$ be a  point, with stabilizer $H=G_x$, whose orbit 
$G x$ is closed.
 Suppose  every 
$H$-module $L$ that appears in $N_{[x]}^*$ is $G$-separable 
(Definition~\ref{dgseparable}).
Then,  for some $G$-invariant  $\beta \in \C[X]$ not vanishing at $x$, 
and  non-$H$-admissible, irreducible  $G$-submodules $P_i \subseteq \C[X]$,
$1\le i \le r$, with $r < \mbox{codim}(G \cdot [x], X)$, 
 $\spec(\C[X]/J)
 \cap  X(\beta) = G \cdot [x] \cap X(\beta)$, where $J$ denotes the 
ideal generated by the $P_i$s.
\end{theorem} 
(Here  we are   identifying a variety with 
the corresponding reduced scheme supported by it.)

\proof 
By Proposition~\ref{peval}, or rather its proof, 
the functions in every non-$H$-admissible $P$ within $\C[X]$ must 
vanish on $G \cdot [x]$. We need to show 
show that, for some $G$-invariant $\beta$ not vanishing at $x$,
the zero set of $J$ within $X(\beta)$ equals $G \cdot [x] \cap X(\beta)$
scheme-theoretically.

\noindent{\bf \'Etale slice theorem (page 198 in \cite{mumford},  
\cite{luna}):}
Let $x$ be a  point of an affine, smooth, irreducible  $G$-variety $X$,
whose orbit $G x \subseteq X$ is closed.
Then  there exists a smooth,  affine  $H$-variety
 $Y \subseteq X$ passing through  $x$ and a strongly \'etale  map $\psi$
from 
$G\times_H Y$ 
 to a  $G$-invariant 
neighbourhood of $G \cdot x$ in $X$ of the form $X(\alpha)$, for some 
 $G$-invariant $\alpha \in \C[X]$.

Here $Z=G \times_H Y$ denotes the induced $G$-equivariant 
 fibre bundle, with base $G/H$ and fibre 
isomorphic to $Y$ \cite{mumford}. 
Strong \'etale-ness of $\psi$ means that  the map $\psi/G$  from
the quotient $Z/G$ to $X/G$ is \'etale and that the induced natural map  from 
$Z$  to $X \times_{X/G} Z/G$,
the $G$-variety 
obtained from $X$ by base extension, is a $G$-isomorphism.

The slice theorem  suggests that we  prove our theorem in two steps.
 First,  consider the case
 when $X$  is a fibre bundle of the form
$G\times_H Y$, where $Y$ is a smooth affine variety and then make a transition
to the general case. Note that $H=G_x$ is reductive since $G x \subseteq X$ is
closed and hence affine \cite{matsushima}.

We shall need the following:

\begin{prop} \label{galbal}
Let $V$ be a finite-dimensional irreducible $G$-module, $G$ connected and
reductive, with
basis co-ordinate functions $V_1 ,\ldots ,V_s $, 
Let $g\in V$ be a point with closed, affine  orbit $G g \subseteq V$.  
Further, let $I(g)$ be the ideal of $G g$.
 Let $J$ be an ideal of $\C [V]$ such that\\
(i) The variety of $J$ is precisely the orbit $O=G g$. \\ 
(ii) The ideal $J$ is itself $G$-invariant. \\
(iii) There are elements $w_1 ,\ldots ,w_k \in J$  such that 
\ignore{for which we
form the co-tangent vectors $W=\{ dw_i = \sum \frac{\partial
w_i }{\partial V_j } dV_j \} $. $W$ is such that $\{ v\in TV_g |
dw_i (g)(v)=0$ for all $i \}= TO_g $. In other words, }
the tangent space $TO_g $ of the orbit $O$ at $g$ consists of precisely the 
tangent vectors in $TV_g $ annihilated by the differential forms $dw_i$'s.\\
Then $J=I(g)$; i.e., $\spec(\C[V]/J) = O$.

Suppose (i) is replaced by the weaker condition:

(i)' The variety of $J$ contains the orbit $O=G\cdot g$.

Then there exists a $G$-invariant neighbourhood $U_g$ of 
the orbit $G g$ such that  the zero set of $J$ restricted 
to $U_g$ coincides with $G g$ scheme-theoretically.
\end{prop}

\noindent
\proof The $G$-invariance of $J$ and the
connectedness of $G$ implies that all associated primes of $J$
must themselves be $G$-invariant. Since there are no proper
$G$-invariant subsets of $O$, we  conclude that there are no
associated primes of $J$ other than $I(g)$. Now (iii)  may
be used to apply the  `Jacobian Criterion' (Matsumura\cite{matsumara},
 Theorem 30.4) locally.
The $G$-invariance of $J$ shows that (iii) holds at every point
$y \in O$. The global assertion then follows. \qed

Given an $H$-module $M$, we denote by $\C[M]$
the $H$-module $\sum_{i\ge 0} \sym^i(M^*)$, i.e., the space of 
polynomial functions on $M$. Let $N$ denote the tangent space to $Y$ at 
$x$; it  is an $H$-module. Now we  prove the theorem for 
the variety  $G \times_H N$.

\begin{lemma} \label{lfirststep}
The theorem holds when $X=G\times_H N$ and 
  $x = (1_G, 0_N)$  is the base  point 
on its null section $G/H$,  with  stabilizer $G_x=H$.
\end{lemma}
\proof 
In this case $N$ can be identified with the
 normal space $N_x$ at $x$ to the orbit $G \cdot x = G/H$.
Let $N=\sum_R R$ be an $H$-module decomposition of $N$ into irreducibles. 
Then we can write $N_x = N_{[x]} + M_x$, where 
$\bar N=N_{[x]}$ is the sum  of all nontrivial $H$-submodules $R$ in this
decomposition, and $M_x$ is the sum of all trivial $H$-submodules.
The subvariety $G \cdot [x]= G \cdot M_x$, and 
the codimension of $G \cdot [x]$ is just the dimension 
of $N_{[x]}$.

For any $H$-submodule $L$ of $N_x$, 
consider the induced 
bundle $F(L^*): G\times_H L^*\rightarrow G/H$. Let ${\cal O}_{F(L^*)}$
 be the sheaf of
 germs of sections of this bundle. Let $H^0(G/H, {\cal O}_{F(L^*)})$
 be the $G$-module of 
its global sections. These global sections are regular functions on 
$G\times_H L$ that are linear on each fibre. 
Clearly  $H^0(G/H, {\cal O}_{F(N_{[x]}^*)})$ is a $G$-submodule of 
 $H^0(G/H, {\cal O}_{F(N_x^*)})$, whose elements are regular functions on 
$X$ linear on each fibre. 
Since $G$ is connected, we can apply
 Jacobi's criterion (Proposition~\ref{galbal}) and 
the transitivity of $G$-action. Hence
 it suffices to show that 
the   sections in the
 non-$H$-admissible $G$-submodules of 
$H^0(G/H, {\cal O}_{F(N_{[x]}^*)})$, when restricted to the fibre
$N_{[x]}^*$ at $x$, span $N_{[x]}^*$; clearly the number  $r$ of 
such submodules is less than $\dim(N_{[x]})=\codim(G \cdot [x], X)$.

Let $N_{[x]}= \oplus_R R$ be an $H$-module direct sum 
decomposition of $N_{[x]}$, where 
each $R$ is an irreducible, nontrivial $H$-submodule. Then 
$N_{[x]}^* = \oplus_R R^*$, as an $H$-module, so we get
 a  natural 
$G$-module decomposition
\[ H^0(G/H, {\cal O}_{F(N_{[x]}^*)}) = \oplus_R H^0(G/H, {\cal O}_{F(R^*)}).\]
Hence, it suffices to show that for each $R$ in this decomposition,
there exists a  non-$H$-admissible $G$-submodule of 
$H^0(G/H, {\cal O}_{F(R^*)})$, whose sections,
 when restricted to the fibre 
$R^*$ at $x$, span $R^*$.

So, let $R$ be any such 
nontrivial, irreducible 
  $H$-submodule in this decomposition and $L=R^*$ its dual.
By the Peter-Weyl theorem (eq.(\ref{eqpeterwe}))
\begin{equation} \label{epweyl3}
H^0(G/H,{\cal O}_{F(L)}) = \oplus_Q Q \otimes Hom(Q, L)^H,
\end{equation}
where $Q$ ranges over all finite dimensional irreducible $G$-modules, and 
$Hom(Q,L)^H$ denotes the vector  space of $H$-equivariant linear 
maps from $Q$ to $L$. Thus the $G$-modules $Q$ that appear in 
$H^0(G/H,{\cal O}_F)$ are precisely the ones that contain $L$.
By our $G$-separability assumption, there exists a nonadmissible, irreducible
$G$-module $Q_L$ containing $L$. By eq.(\ref{epweyl3}),
 $H^0(G/H,{\cal O}_{F(L)})$ 
contains a copy of  $Q_L$. Fix one such copy; we denote it
 by $Q_L$ again. The restriction of $Q_L$ to the fibre  $L$ of $F(L)$ at $x$ 
is precisely $L$. Hence the basis elements of $Q_L$ when restricted 
to $L$ span $L$.
 \qed

For every $R$ that appears in the $H$-module decomposition of $N_{[x]}$,
 let $Q_L \subseteq H^0(G/H, {\cal O}_{F(N_{[x]}^*)})$, $L=R^*$,  
be a fixed copy as in the proof above. Let $\Phi$ be the set of 
such finitely many $Q_L$s, each a non-$H$-admissible, irreducible 
 $G$-module of 
regular functions on $G \times_H N$. The  number $r$  of  $Q_L$'s in
$\Phi$ is less than $\codim(G \cdot [x],X)$.
Since many $R$'s in the $H$-module decomposition may be isomorphic,
many $Q_L$s in $\Phi$  may be  isomorphic as $G$-modules.
The proof above  shows that:

\begin{lemma} \label{eqspan1}
The differentials of the 
 basis elements of the non-$H$-admissible $G$-modules $Q_L$ in $\Phi$,
when restricted to  $N_{[x]}$, 
 span the whole of $N_{[x]}$, and the zero set of the ideal generated 
by them coincides with $G \cdot [x] = G \cdot M_x$ scheme-theoretically.
The  number $r$ of modules in $\Phi$ is 
less than $\mbox{codim}(G \cdot [x], G \times_H N)$.
\end{lemma}

\ignore{
\begin{lemma} \label{eqspan}
The theorem holds when $X= G \times_H Y$, $Y$ a smooth, affine $H$-variety,
and $x \in Y$ a point on the null section $G/H$ of this fibre bundle 
with stabilizer $H$. There exists a  set $\Phi$ of 
 at most $\codim(G/H, X)$ non-$H$-admissible
irreducible $G$-submodules of $\C[X]$, such that 
the differentials of the 
 basis elements of the $G$-modules $Q_L$ in $\Phi$,
when restricted to  $N_{[x]}$, 
 span the whole of $N_{[x]}$. Moreover, there exists a saturated,
$H$-invariant affine neighbourhood $Y' \subseteq Y$ containing $x$ such that
 the zero set (scheme) of the ideal generated 
by these basis elements within $X'=G \times_H Y'$ 
 coincides with $G \cdot [x] \cap  X'$.
\end{lemma}
\proof This easily follows from Lemma~\ref{eqspan1} 
because the slice $Y$ in the statement of the slice theorem
can be chosen so that  there is  a strongly \'etale $H$-morphism 
from $Y$ onto
 an $H$-invariant affine neighbourhood of $x$ in $N=N_x$ (\cite{mumford},
page 198). This gives rise to a strongly \'etale $G$-morphism  $\phi$ from 
$G \times_H Y$ to $G \times_H N$. We have the following diagram: 
\begin{equation} \label{eqdiag}
G \times_H N \stackrel {\phi} \leftarrow G \times_H Y \stackrel \psi 
\rightarrow  X,
\end{equation} 
where the arrows denote strongly \'etale $G$-morphisms.
 We simply let $Q_L$s here be the inverse images, via $\phi$, of the 
$Q_L$s constructed in Lemma~\ref{eqspan1}. The existence of $Y'$ also
follows from Lemma~\ref{eqspan1} in conjunction with strong \'etale-ness.
\qed
}

Now we turn to the general case. 
By the \'etale slice theorem, 
there exists an   affine $H$-variety
 $Y \subseteq X$ passing through   $x$ and a strongly \'etale  map $\psi$
from 
$G\times_H Y$ 
 to a  $G$-invariant 
neighbourhood of $G \cdot x$ in $X$ of the form $X(\alpha)$, for some 
 $G$-invariant $\alpha \in \C[X]$ not vanishing at $x$.
 Since $\C[X(\alpha)] = \C[X]_\alpha$, the 
ideal generated by non-$H$-admissible, irreducible  $G$-submodules of $\C[X]$ 
within $\C[X(\alpha)]$ coincides with the one generated by 
 non-$H$-admissible, irreducible  $G$-submodules of $\C[X(\alpha)]$. 
Hence, in the statement of the theorem,
 we can replace $X$ by $X(\alpha)$.
\ignore{
For the same reason (cf. diagram \ref{eqdiag}) we can replace $Y$ with $Y'$ 
in the statement of Lemma~\ref{eqspan}.
Thus we may assume that there is a strongly \'etale map $\psi$ from
$Z=G\times_H Y$ to $X$, such that the zero set of the ideal generated by 
$Q_L$s in Lemma~\ref{eqspan} in $G \times_H Y$ coincides with $G \cdot 
[x] \subseteq G \times_H Y$.} 
Strong \'etale-ness of $\psi$ implies  \cite{mumford} that
there is an analytic 
  neighbourhood $Y_{an} \subseteq  N_x$ of $x$ in $N_x$--called 
analytic slice through $x$--such that
 $G\times_H Y_{an}$ is $G$-isomorphic to an analytic 
 $G$-invariant neighbourhood
$U$ of the orbit of $x$. However, there may not be an algebraic slice
with this  property, and this forces us in the analytic category in what
follows. 
Since $U \simeq G\times_H Y_{an} \subseteq G\times_H N$, 
each $Q_L$ corresponds to, and can be identified with,
a $G$-module $Q_L(U)$  of analytic functions
on $U$.
By Lemma~\ref{eqspan1}, the zero set of the $Q_L(U)$'s in $\Phi$ 
within $U$ coincides, as a complex space \cite{stein}, with
$G \cdot [x] \cap U$.
 Our goal is to show that
each $Q_L(U)$ can be approximated very closely 
within $U$ by an isomorphic  $G$-submodule of $\C[X]$.
For this, we shall need the following results from  complex function theory.

\noindent  {\bf Cartan-Oka Theorem: (\cite{stein})} 
Let $A$ be a Stein space, and $B$ its closed analytic subspace. Then 
every holomorphic function on $B$ extends to a holomorphic function on 
$A$.

Let ${\cal O}_A$, ${\cal O}_B$ be the sheaves of germs of holomorphic functions
on $A$ and $B$ respectively, and ${\cal I}_B$ the sheaf of ideals of $B$. 
Then by Oka's theorem ${\cal I}_B$ is coherent, and since $A$ is Stein,
its higher cohomology $H^i(A,{\cal I}_B)$, $i>0$, vanishes (Cartan's theorem
B). Hence, this result follows from the long exact cohomology 
sequence associated 
with the exact sequence of sheaves
\[ 0 \rightarrow {\cal I}_B \rightarrow {\cal O}_A \rightarrow {\cal O}_B 
\rightarrow 0,\]
where we consider ${\cal O}_B$ as a sheaf on $A$ via extension by zero.

We shall denote the ring of holomorphic functions on an analytic variety 
$W$ by ${\cal O}(W)$.
\begin{lemma} \label{lcartanoka}
Let $A$ be a Stein $G$-space, and $B$ its closed analytic $G$-subspace,
$G$ a connected 
reductive group. Let 
$M$ be a finite dimensional $G$-submodule of ${\cal O}(B)$. Then 
there exists a $G$-equivariant extension map $\phi: M \rightarrow {\cal O}(A)$.
\end{lemma}
Here, we say that $\phi$ is an  extension map if,
 for any $s \in M$, the restriction of $\phi(s)$ to $B$ coincides
with $s$.

\proof Fix a basis $s_1,\ldots, s_l$ of $M$. By the Cartan-Oka theorem,
each $s_i$ can be extended to a holomorphic  function $\hat s_i$ on 
$A$. Let $\rho: M \rightarrow {\cal O}(A)$ be a linear map defined 
by setting 
$\rho(\sum_i b_i s_i) = \sum_i b_i \hat s_i$. Though $\rho$  need not be 
$G$-equivariant,  it may be converted into a $G$-equivariant 
map by Weyl's unitary trick \cite{fulton}. Specifically,
regard $Hom(M, {\cal O}(A))$ as a $G$-module in the natural way. 
Fix a maximal compact subgroup $E \subseteq G$. Let $d e$ denote left-invariant
Haar measure on $E$, and  let 
\[ \phi = \int_E e(\rho) d e.\]
Then $\phi$ is an $E$-equivariant extension. Since $M$ is finite dimensional,
it follows from the unitary trick that $\phi$ is 
$G$-equivariant  as well.
\qed

\begin{lemma} \label{lpowerexp}
Let $W$ be a linear representation of connected, reductive $G$,
$V$ a linear space with trivial $G$ action, $D \subseteq V$ a ball
around the origin. 
 Let $A=W \times D$.
Then:

\noindent (1) Any holomorphic function
 $a$ on $A$ has a unique power series expansion 
of the form
\begin{equation}
a(w,v)= \sum_{i, j} a_{i}^ {j} w^{j} v^{i}.
\end{equation}
Here $i=(i_1,\cdots,i_r)$,  $r=\dim(V)$, and 
$j=(j_1,\cdots,j_q)$,  $q = \dim(W)$, are tuples of nonnegative 
integers; and  $v^{i}= v_1^{i_1}\cdots v_r^{i_r}$, and 
$w^{j}=w_1^{j_1}\cdots w_q^{j_q}$, where 
 $v_1,\ldots, v_r$ are   the coordinates of $V$, and 
$w_1,\ldots, w_q$ are   the coordinates of $W$.

\noindent (2) For any $k=(i, d)$, the map $\delta_k: {\cal O}(A) \rightarrow 
\C[W \times V]= \C [W] \otimes \C[V]$, which maps 
$a$ to $\sum_{j: j_1+\cdots + j_q= d} a_i^j w^j v^i$ is $G$-equivariant.
\end{lemma}
\proof  The first statement follows because $A$ is a proper Reinhardt domain
in $W \times V$ (cf. \cite{grauertf}, page 20).
\ignore{Let $p$ denote the 
projection of $A$ onto its second factor $D$. 
 For each $v \in D$, the restriction of $a$ to the fibre 
$p^{-1}v \cong W$
is
holomorphic on the whole of $W$. Hence, 
for each  $v \in D$,
 $a (w,v)$ has a unique power series expansion of 
the form 
\begin{equation} 
a(w,v)  = \sum_{i \ge 0}
a_{i}(v) w^i,
\end{equation} 
which converges over the whole of $W$ (\cite{grauertf}, page 20).
Furthermore, since $a$ is holomorphic on $A$, it also 
follows, by elementary complex variable theory, that each 
coefficient $a_i(v)$ is holomorphic on $D$. Hence, it too 
admits a unique power series expansion 
\[ a_i(v) = \sum_{j \ge 0} a_i^j v^j,\]
that converges in $D$ (\cite{grauertf}, page 20). This proves the 
first statement.}

Each $w_1^{j_1}\cdots w_q^{j_q}$ is a polynomial (regular) function 
on $W$ and is contained in the finite dimensional $G$-submodule in $\C[W]$ 
of homogeneous forms of degree $d=j_1+ \cdots + j_q$. Since the $G$-action on
$D$ is trivial, it follows that each $\delta_k$ is $G$-equivariant.
\qed

Let $X$ be an affine, smooth $G$-variety, $G$ a connected reductive group. Let 
$\psi$ be the projection from $X$ to its quotient $X/G$. Let 
$x$ be a point in $X$ with closed orbit $G x \subseteq X$,
 and $\bar x = \psi(x)$ its projection. 
Embed the affine variety $X/G$ in a linear space $V$, with $\bar x$ at its
origin. Suppose 
$U_{\bar x}$ is a Stein neighbourhood of $\bar x$ in $X/G$,
such that $U_{\bar x} = D \cap X/G$, where $D \subseteq V$ is a ball around 
$\bar x$. Let 
$U=\psi^{-1} (U_{\bar x})$. 
\begin{lemma} \label{lcrucial}
Let 
$Q$ be a finite dimensional $G$-submodule of ${\cal O}(U)$. 
Then there exist $G$-equivariant linear maps $\rho_k: Q \rightarrow \C[X]$
such that any  $s \in Q$ admits a power series expansion 
 $s = \sum_{k=0}^\infty \rho_k(s)$ that converges  everywhere in 
$U$.
\end{lemma} 
\proof 
We can embed $X$ $G$-equivariantly as a closed affine $G$-subvariety of 
some  linear representation $W$ of $G$ (\cite{kempf}, Lemma 1.1).
Let $A= W \times D$, which is Stein.
 It has a $G$-action, the action on $D$ being trivial.
 Let $B \subseteq A$ be the  closed analytic $G$-subspace 
consisting points $(x,u)$, with $x \in X$, $u \in U_{\bar x}$,
and  $\psi(x)=u$. 
It is   isomorphic to $U$. So $s$ corresponds to a holomorphic function
on $B$, which we shall denote by $s$ again. Thus 
we can regard $Q \subseteq {\cal O}(B)$.

By Lemma~\ref{lcartanoka}, there exists a $G$-equivariant extension
map $\phi: Q \rightarrow {\cal O}(A)$. Let $\delta_k$ be the
$G$-equivariant map of Lemma~\ref{lpowerexp} applied to $A$.
Finally, let $\alpha:\C[W \times V] \rightarrow \C[X]$
 be the $G$-equivariant restriction map corresponding to the 
$G$-equivariant embedding $X  \rightarrow W \times V$,
which maps $x \in X$ to $(x, \psi(x))$.
 Let $\rho_k = \alpha \circ \delta_k \circ 
\phi$. Then $s \in Q$ has a $G$-equivariant power series expansion 
\[ s = \sum_k \rho_k(s),\]
that converges everywhere in $U$. \qed

Now we return to the proof of Theorem~\ref{tnonadmrel}.
Let $\psi$ be the strongly \'etale map from $G \times_H Y_{an}$ 
to a $G$-invariant neighbourhood $U$ of the orbit $G\cdot x$. Here
$Y_{an} \subseteq N_x$ is an analytic slice, and $U$ is of the form
$\psi^{-1} U_{\bar x}$, where $U_{\bar x}$
is an analytic neighbourhood 
of $\bar x = \psi(x)$. We can assume that $U_{\bar x}$ is Stein, of
the form $D \cap X/G$ as in Lemma~\ref{lcrucial},
 for a small enough ball $D$ around
$\bar x$ in $V \supseteq X/G$. 
Let $Q_L \in \Phi$
 be the finitely many, irreducible, non-$H$-admissible $G$-submodules 
 of the ring of regular functions on 
$G\times_H N$  as  in Lemma~\ref{eqspan1}; their number $r$ is less than
$\codim(G\cdot [x],G\times_H N) = \codim(G\cdot [x],X)$.
  We shall  denote the  restriction of $Q_L$ to
$G\times_H Y_{an}$  by $Q_L$ again. It corresponds  to a $G$-module of 
analytic functions on $U$, which we shall denote by $Q_L(U)$;
the analytic functions in $Q_L(U)$ though may not extend to the whole 
of  $X$.

Now we come to the crux of the proof. The $G$-module 
 $Q_L(U)$ is isomorphic
to $Q_L$,  and hence,  finite dimensional. Hence
we may apply  Lemma~\ref{lcrucial}. 
Let $\rho_k(L)$ denote the $G$-equivariant 
 projection from $Q_L(U)$ to $\C[X]$ therein. 
Let $\tilde \rho_k(L)= \sum_{j \le k} \rho_j(L)$.  When
$k$ is large enough $\tilde \rho_k(L) (Q_L(U))$ will be 
a good approximation to $Q_L(U)$. 
Let $Q_L^k \subseteq \C[X]$ be the $G$-module  that is the image of 
this 
$G$-equivariant projection $\tilde \rho_k$. Since 
$Q_L(U) \simeq Q_L$
 is irreducible,   $Q_L^k$ is either 
zero, or isomorphic to $Q_L$. When $k$ is large enough,
 $Q_L^k$ is 
 isomorphic to $Q_L$--hence it is
 non-$H$-admissible, and vanishes on $G \cdot [x]$.

Since 
$U$ is $G$-isomorphic to $G \times_H Y_{an} \subseteq G \times_H N$, 
it follows from Lemma~\ref{eqspan1} that 
the 
 the differentials of the  basis functions in all the $Q_L(U)$ in $\Phi$,
 when 
restricted to $N_{[x]}$, 
 span the whole of $N_{[x]}^*$.
We  approximate each 
$Q_L(U)$  by 
$Q_L^k \subseteq \C[X]$ for a large enough $k$.
When $k$ is large enough, 
the differentials of the basis functions in $Q_L^k$,
when restricted to $N_{[x]}$, will also span the whole of $N_{[x]}^*$.
But each $Q_L^k$ is a non-$H$-admissible, irreducible
$G$-submodule of $\C[X]$. Thus it follows that 
the differentials of the   basis functions of 
the  non-$H$-admissible, irreducible 
 $G$-submodules $Q_L^k$ 
of $\C[X]$, for $k$ large enough,  span $N_{[x]}^*$.
 Because of the transitivity of the 
$G$-action, the same holds for all points in
the orbit of $x$.
Since $G$ is connected, and  all $Q_L$'s are $G$-modules, it now 
 follows from
 the Jacobian criterion (proposition~\ref{galbal}, or rather its proof)
and the fact that $U \simeq G\times_H Y_{an}$,
that the zero-set 
 of the basis functions of these $Q_L^k$'s within $U$ coincides with 
$G \cdot [x] \cap U$ scheme theoretically (i.e., as a complex space 
\cite{stein}).
Since $Q_L^k$s are $G$-submodules of $\C[X]$, there exists a Zariski-open 
$G$-invariant neighbourhood $U' \supseteq U$ such that the zero set of
$Q_L^k$s within $U'$ coincides with $G \cdot [x] \cap U'$ 
scheme theoretically.
It remains to show that $U'$ can be chosen to be of the form 
$X(\beta)$, for some $G$-invariant $\beta$. The projection $\psi(U')$ 
into $X/G$ is a constructible \cite{hartshorne} set that 
contains $U_{\bar x}$. Hence $\psi(U')$ contains
 a Zariski-open 
affine neighbourhood of the form $(X/G)_\alpha$ for some $G$-invariant 
$\alpha$ not vanishing at $x$. Its inverse $\psi^{-1}(X/G)_\alpha$ 
is of the form $X(\alpha)$ and has the required properties. \qed

\noindent{\bf Remark}: Suppose every 
$H$-module  that appears in $N_{[x]}^*$ is not $G$-separable,
as assumed in  Theorem~\ref{tnonadmrel}. Then
one can similarly prove a weaker assertion that
  for some $G$-invariant analytic neighbourhood $U$ (as in the proof above) 
 of 
$G x$, $\spec(\C[X]/J) \cap U$, as a complex space \cite{remmertcoh}, is a 
 subspace of $G\times_H \spec(I)$, where $\spec(I)$ is a 
subscheme of $N_{[x]}$ and 
$I \subseteq \C[N_{[x]}]$ is the ideal generated by
the  $G$-separable $H$-submodules of 
 $\C[N_{[x]}]$.

\section{Partial stability}  \label{spartiallystable}
Let $V$ be a linear representation of $G$.
Let $P=KU$ be a parabolic subgroup of $G$, and 
 $R$ a reductive subgroup of $K$.

\begin{defn} \label{dpartiallystable}
We say that $v \in P(V)$ is $(R,P)$-stable
({\em partially stable}) if (1) it is stable with respect to
the restricted action of $R$ on $V$, and (2) $U \subseteq G_v
 \subseteq P$.
\end{defn}
Here $U \subseteq G_v$ implies that $U \subseteq G_{\hat v}$.
The defect $\delta(v)$ of $v$ is defined to be the difference between
the ranks of the root systems of $R$ and $K$. 
In our applications, the defect will be  small--in fact, just one--and 
 $R$ will always be a 
semisimple Levi subgroup of
a parabolic subgroup of $K$--so that the root system of $R$ will always be 
a subsystem of that of $K$. 

A stable point  of $V$ is $(G,G)$-stable.
A point
 $v \in P(V)$ is  $(R,P)$-stable iff it is an
$(R,K)$-stable point of $P(Y)$, where 
 $Y=V^U$ is the $K$-module  of $U$-invariants in $V$.

\noindent {\bf Example 1:} 
The simplest example of a partially stable point with defect zero 
is the point $v=v_\lambda  \in P(V)$ that corresponds to
 the highest weight vector of an irreducible $G$-representation
$V=V_\lambda(G)$.  The stabilizer $P=G_v$ is parabolic, and $v$ is clearly
$(L,P)$-stable, where $L$ is a semisimple Levi subgroup of $P$.

\noindent {\bf Example 2:} 
let $f=\phi(h)$ be as in Definition~\ref{dpstablebasic}, with $h$ stable. Then
$f$ is $(R,P)$-stable, with defect one, with respect to the
action of $G$ (as well as $\hat G$), where:
 $P$ is a  parabolic subgroup of $G$ (resp. $\hat G$), whose 
elements  transforms the variables in $\bar X$ to
their linear combinations, thus  preserving
an appropriate  flag $\C^{k+1} \subseteq \C^l$, and $R \simeq SL_k(\C)
 \times SL_{l-k-1}$ is   naturally embedded in 
the semisimple Levi subgroup of $P$ isomorphic to 
$SL_{k+1}(\C) \times SL_{l-k-1}(\C)$.

\begin{defn} \label{drestr} 
Given dominant weights $\alpha$ and $\beta$ of $R$ and $K$,
we shall say that $\alpha \lhd_R^K \beta$, or $\beta \rhd_R^K \alpha$,
if $V_\alpha(R)$ occurs in  $V_{\beta}(K)$, dropping the superscript
or subscript whenever possible. 
\end{defn} 

In the definition of $(R,P)$-stability the group $R$ will usually be such that 
\begin{equation} \label{eqrestrictr}
\tilde L \subseteq R \subseteq \tilde K \subseteq K,
\end{equation} 
 for some
parabolic subgroup $\tilde P=\tilde T \tilde L \tilde U = \tilde K \tilde U$
of $K$, as in Example 2. 
Then, using Littelmann's restriction rule \cite{littelmann}, 
one can determine how any irreducible representation $V_\beta(K)$
explicitly decomposes as a $\tilde K$-module (and hence as an $R$-module).
This, in turn, gives 
an explicit relationship between
 $\alpha$ and $\beta$ in Definition~\ref{drestr}.

In Example 2 above, 
$K \simeq
GL_{1+k}(\C) \times GL_{l-1-k}$ 
and $R \simeq  SL_k(\C) \times SL_{l-1-k}(\C)$. 
In this case,  Littelmann's restriction rule  reduces to a variant 
of the well known Pieri's branching rule \cite{fulton},
which gives an explicit 
decomposition of $V_\mu(GL_{1+k}(\C))$ as a 
$GL_k(\C)$ module.

For a connected reductive group $D$, we shall denote by 
$i_D$ the canonical involution of its dominant weights so that 
$V_\lambda(D)^*=V_{i_D\lambda}(D)$.
Let $v\in P(V)$ be an  $(R,P)$-stable 
 point as above.
Let $W$ and $Y$ be respectively 
the smallest $K$-submodule
and $R$-submodule of $V$ containing $\hat v$.

\begin{defn} \label{dlieover}
We  say that 
 a dominant weight $\beta$ of $G$ lies over 
a weight $\mu$ of $R$ at $v$ and degree $d$
if 
\begin{enumerate} 
\item  $V_\mu(R)$ and $V_{\beta'}(K)$ occur in
$R_Y[v]_d$ and $R_{W}[v]_d$ respectively, where 
$\beta' = i_{K} (i_G \beta)$, and
\item  $\mu \lhd_R^K \beta'$. 
\end{enumerate}
We say that  a dominant weight $\beta$ of $G$ lies over 
a weight $\mu$ of $R$ at $v$ if this is so at some $d$.
\end{defn}

This definition does not depend on
 the choice of a Levi subgroup $K \supseteq R$ of $P$,
because  $U \subseteq G_v$. 
When the defect is zero, and $R$ satisfies eq.(\ref{eqrestrictr}),
2. just says that the weight $\beta'$, restricted to $R$, is equal to $\mu$.
The number of $\beta$ lying over $\mu$ at a fixed $d$ depends on the defect;
it is small if 
 the defect is small.

\section{Borel-Weil for a partially stable point} \label{subspstable}
In this section we shall prove 
Theorem~\ref{tlieovernew} (b) for   partially stable points. 
Its precise statement is as follows.

\begin{theorem} \label{tlieoverpstable} 
Suppose $v$ is  $(R,P)$-stable (cf. Definition~\ref{dpartiallystable}). Then
$V_\lambda(G)$ can occur in $R_V[v]$ only if
$\lambda$  lies over some $R_{\hat v}$-admissible dominant weight $\mu$  of $R$
at $v$
(cf. Definition~\ref{dlieover}).
Conversely,
 for every $R_{\hat v}$-admissible dominant weight $\mu$ of $R$,
$R_V[v]$ contains $V_\lambda(G)$ for some
 dominant weight $\lambda$ of $G$ 
lying over $\mu$ at $v$.
\end{theorem}

This will follow 
from the following stronger result. 

Suppose $v \in P(V)$ is partially stable,
specifically $(R,P)$-stable, where 
$P=TLU=KU$, and $R \subseteq K$. 

\ignore{and $\tilde P=\tilde T \tilde K \tilde U$. 
 Given  dominant weights $\mu$ and $\beta$  of $\tilde K$ and $K$,
 we shall say that
 $\mu \lhd_{underline K}^K \beta$, or equivalently, that 
$\beta \rhd_{underline K}^K \mu$, if
$V_\mu(\tilde K)$
 occurs in a complete $\tilde K$-module decomposition of 
$V_\beta(K)$. This means that $\beta$ and $\mu$ are related 
as per the rule above.
If $\tilde K$ and $K$
 are clear from the context, we shall drop the superscript
and the subscript.}

Let $W$ and $Y$ be respectively 
the smallest $K$-submodule,
$R$-submodule  of $V$ containing $\hat v$. Let ${\cal O}(d)$ be the 
twisting sheaf on $P(V)$ and ${\cal O}_{\Delta_V[v]}(d)$, 
${\cal O}_{\Delta_W[v]}(d)$ the corresponding invertible sheaves on
 $\Delta_V[v]$ and $\Delta_W[v]$ respectively.
Let \\
$\Gamma(\Delta_V[v],{\cal O}_{\Delta_V[v]}(d))$ 
and $\Gamma(\Delta_W[v],{\cal O}_{\Delta_W[v]}(d))$ be the 
$G$ and $K$-modules of their   of global sections.
Clearly $R_V[v]_d \subseteq \Gamma(\Delta_V[v],{\cal O}_{\Delta_V[v]}(d))$ 
for all $d \ge 0$. We have an  equality for all $d \ge 0$ iff
$\Delta_V[v]$ is  projectively normal.
(cf. page 126, Hartshorne \cite{hartshorne}).
Similarly, 
 $R_W[v]_d \subseteq \Gamma(\Delta_W[v],{\cal O}_{\Delta_W[v]}(d))$ for 
all $d \ge 0$, with equality if $\Delta_W[v]$ is  projectively normal.

The following result shows that the $G$-module structure of $R_V[v]$ is
ultimately related to the $R$-module structure of $R_Y[v]$. In turn,
 we already 
know which $R$-modules can occur in $R_Y[v]$ since $v \in P(Y)$ is stable
with respect to the action of $R$ (Theorem~\ref{tlieovernew} (a)).

\begin{theorem} \label{tlieoverrefined}
({\em Borel-Weil for partially stable points}) 

Suppose $v \in P(V)$ is
 $(R,P)$-stable as above. Then
\begin{enumerate} 
\item The $G$-module structure of $R_V[v]$
is equivalent to  the $K$-module structure of 
$R_{W}[v]$: Specifically, the multiplicity of 
 a  $G$-module $V_\lambda(G)$ in $R_V[v]_d^*$ is equal to the 
multiplicity of the $K$-module $V_\lambda(K)$ in
$R_{W}[v]_d^*$, where $\lambda$ is regarded as a dominant weight of 
$K$ by restriction. Moreover, a
$K$-module $V_\alpha(K)$ can occur in 
$R_{W}[v]_d^*$ only if $\alpha$ is also a  dominant weight of $G$.

\item The multiplicity of  $V_\lambda(G)$ in the module 
$\Gamma(\Delta_V[v],{\cal O}_{\Delta_V[v]}(d))^*$ of global sections
of ${\cal O}_{\Delta_V[v]}(d)$
 is less than or equal to
the multiplicity of $V_\lambda(K)$ in 
$\Gamma(\Delta_W[v],{\cal O}_{\Delta_W[v]}(d))^*$. 
If $\Delta_W[v]$ is projectively normal  then the 
 two multiplicities are
equal, for all $\lambda$ and $d \ge 0$,
  and $\Delta_V[v]$ is also projectively normal. 

\item A   $K$-module $V_\beta(K)$ can occur in
$R_{W}[v]_d$ only if for some dominant weight 
$\alpha \lhd_{R}^{K} \beta$
 of $R$, 
 $V_\alpha(R)$ occurs in $R_{Y}[v]_d$. Conversely,
for every $R$-module 
$V_\alpha(R)$ occurring in $R_{Y}[v]_d$, there exists a 
dominant weight $\beta \rhd^{K}_{R} \alpha$ of $K$ such that 
$V_\beta(K)$  occurs in
$R_{W}[v]_d$.

\item Finally, an $R$-module $V_\mu(R)$ occurs in $R_Y[v]$--i.e. in some $R_Y[v]_d$--iff it is 
$R_{\hat v}$-admissible.
\ignore{$V_\alpha(K)$, 
is $K_{\hat v}$-admissible,
and hence, also  $L_{\hat v}$ admissible. Conversely,
every irreducible  $L_{\hat v}$-admissible $L$-module occurs in 
$R_{W}[v]$, i.e. in some $R_{W}[v]_d$.}
\end{enumerate}
\end{theorem}

\noindent {\bf Remark 1} In the third
statement, it is desirable 
that we have an explicit criterion for deciding if $\alpha \lhd_R^K \beta$.
When $R$ satisfies eq.(\ref{eqrestrictr}) in 
Section~\ref{spartiallystable}, such a criterion is given by 
Littlemann's rule as pointed out  there.

\noindent {\bf Remark 2:} When $G$ is semisimple and simply connected,
 and 
$v$ corresponds to the highest weight vector 
in $V=V_\lambda(G)$,
 $\Delta_V[v]=G/P$, and $\Delta_W[v]$ is just the point $v$.
Hence $\Gamma(\Delta_W[v],{\cal O}_{\Delta_W[v]}(d))^*=V_{d \lambda}(K)$, for 
$d \ge 0$.
The second statement now implies that 
 $\Gamma(\Delta_V[v],{\cal O}_{\Delta_V[v]}(d))^*=\Gamma(G/P,
{\cal O}_{G/P}(d))^*= V_{d \lambda}(G)$, for $d \ge 0$--which 
is  the Borel-Weil theorem
\cite{knapp}. 

We will first prove two propositions. For that
we  need the following lemma from representation theory.

\begin{lemma} \label{lknapp} (cf. Theorem 5.104 \cite{knapp})
Let $V_\lambda(G)$ be an irreducible representation of a connected reductive
group $G$
with highest weight $\lambda$. Let $P =KU \subseteq G$ be a parabolic subgroup.
 Then $V_\lambda(G)^{U} = V_\lambda(K)$; here 
 $V_\lambda(G)^{U}$ is the subspace of $U$-invariants in
 $V_\lambda(G)$.
\end{lemma} 
\ignore{Here a  weight of $G$ can be considered a weight of $K$ as well since 
we can assume that the maximal torus with respect to which  the weights are 
defined are the same for $G$ and $K$.}
\ignore{
\proof Let ${\cal G}$, ${\cal L}$ and ${\cal U}$ denote the Lie algebras of 
$G$, $L$ and $U$ respectively. 
The root system of ${\cal L}$ is
a subsystem of the one for ${\cal G}$. Hence, in what follows,
 each simple root of ${\cal L}$ will  also be considered as
a simple root of ${\cal G}$.

Since $L$ is in the normalizer of $U$,
  $V_\lambda(G)^{U}$ is an $L$-module.
 Since the highest weight
vector $v_\lambda$ of $V_\lambda(G)$ is invariant under $U$, it belongs to
$V_\lambda(G)^{U}$, and hence $V_\lambda(L) \subseteq V_\lambda(G)^U$.
Moreover, $V_\lambda(L)$ is its  only $L$-submodule  with weight $\lambda$,
because  the highest weight vector (i.e. the one with weight $\lambda$) 
 in $V_\lambda(G)$ is unique \cite{fulton}.
Suppose to the contrary that $V_\lambda(G)$  contains an $L$-module 
$V_\alpha(L)$ with $\alpha \not = \lambda$. 

Let $w_\alpha$ be its highest vector; i.e., the vector that is
killed by the positive simple roots of ${\cal L}$, the Lie algebra of $L$.
 If it is also invariant under
$U$, then it is also killed by all positive 
simple roots in ${\cal U}$, the Lie algebra
of $U$. In other words, it is killed by all positive 
simple roots of ${\cal G}$ 

Hence it must be $v_\lambda$, since the highest weight vector in $V_\lambda(G)$
is unique. Thus $\alpha = \lambda$. \qed
}

Let $z \in P(V)$ be a point whose stabilizer $G_z \subseteq G$ contains
$U$, so that
the stabilizer $G_{\hat z} \subseteq G$ of $\hat z \in V$ also
 contains $U$. 
Let $Z$ be the smallest $K$-submodule of
 $V$ containing  $\hat z$. Let $i$ denote the 
embedding of $Z$ in $V$. 
 The following
result shows that  $R_{Z}[z]$ and  $R_V[z]$ are 
closely related.

\begin{prop} \label{plift}
(a) The multiplicity of an irreducible  module $V_\lambda(G)$  
in $R_V[z]_d^*$ is equal to the multiplicity of 
 $V_\lambda(K)$
 in $R_{Z}[z]_d^*$.  Moreover, 
 $V_\alpha(K)$ can occur in 
$R_{Z}[z]_d^*$ only if $\alpha$ is also a  dominant weight of $G$.

(b) The multiplicity of  $V_\lambda(G)$ in 
$\Gamma(\Delta_V[z],{\cal O}_{\Delta_V[z]}(d))^*$ is less than or equal to
the multiplicity of $V_\lambda(K)$ in 
$\Gamma(\Delta_Z[z],{\cal O}_{\Delta_Z[z]}(d))^*$.
If $\Delta_Z[z]$ is projectively normal  then the 
 two multiplicities are
equal for all $\lambda$ and $d \ge 0$, and $\Delta_V[z]$ is also projectively normal. 
\end{prop} 
\proof 
 Since the stabilizer $G_{\hat z}$ contains $U$, and $U$
 is normalized by $K$,
the stabilizer of every point in $Z$ contains $U$; in other words, the action
of $U$ on $Z$ is trivial. Thus $Z$ can be considered a $P$-module.
 The embedding map $i: Z \rightarrow V$ is then
$P$-equivariant. By restriction, we get a $P$-equivariant, closed 
embedding $i: \check \Delta_Z[z] \rightarrow \check \Delta_V[z]$, where 
$\check \Delta_V[z] \subseteq V$ and $\check \Delta_Z[z] \subseteq Z$ denote 
 the 
affine cones of $\Delta_V[z]$ and $\Delta_Z[z]$.
 Hence, the corresponding surjection $i^*: R_V[z] \rightarrow
R_Z[z]$ is $P$-equivariant.  Since it is degree-preserving,
by restriction, we get a $P$-equivariant surjection $i^*: R_V[z]_d \rightarrow
R_Z[z]_d$  for every $d$.
Since the action of $U$ on $Z$ is trivial, we get the dual
injection $i: (R_Z[z]_d)^* \rightarrow (R_V[z]_d^*)^U$.

Let $M$ be any irreducible $G$-submodule of $R_V[z]_d$.
Not all functions in $M$
 can vanish at $z$--otherwise 
arguing as in the proof of Proposition~\ref{peval},
 we can conclude that 
the functions in $M$  vanish identically on the affine cone
$\check\Delta_V[z]$, which is not possible.
 It follows that the restriction map $i^*$ is nonzero
on $M$.
Thus $N=i^*(M)$ is a nonzero $K$-module, with trivial
$U$-action. Dually, this means $i(N^*)$ is a nonzero $K$-submodule of
$(M^*)^U$. If $M^*=V_\lambda(G)$, then $(M^*)^U = V_\lambda(K)$ 
(Lemma~\ref{lknapp}), and hence irreducible. So $(M^*)^U \simeq i(N^*)$.
Thus the injection $i:(R_Z[z]_d)^* \rightarrow (R_V[z]_d^*)^U$
is an isomorphism. Hence the multiplicity of $V_\lambda(G)$ in 
$R_V[z]_d^*$ is equal to the multiplicity of $V_\lambda(K)$ in 
$R_Z[z]_d^*$, and,
moreover, 
 $V_\alpha(K)$ can occur in 
$R_{Z}[z]_d^*$ only if $\alpha$ is also a  dominant weight of $G$.
 This proves (a) 

The proof of (b) is similar.
 The embedding map $i: Z \rightarrow V$ induces a
 $P$-equivariant map  $i^*: \Gamma(\Delta_V[z],{\cal O}_{\Delta_V[z]}(d))^*
 \rightarrow \Gamma(\Delta_Z[z],{\cal O}_{\Delta_Z[z]}(d))^*$,
 which need not be surjection in general.
Let $M$ be any irreducible $G$-submodule of \\
$\Gamma(\Delta_V[z],{\cal O}_{\Delta_V[z]}(d))$. One shows similarly
that $N=i^*(M)$ is a nonzero $K$-submodule of
$\Gamma(\Delta_Z[z],{\cal O}_{\Delta_Z[z]}(d))$, with trivial
$U$-action, and $(M^*)^U \simeq i(N^*)$. This proves the first statement
of (b).

If $\Delta_Z[z]$ is projectively normal, i.e., its homogeneous 
coordinate ring is integrally closed, then  
$\Gamma(\Delta_Z[z],{\cal O}_{\Delta_Z[z]}(d)) = R_Z[z]_d$, for $d \ge 0$
 (cf. page 126 
Hartshorne \cite{hartshorne}).  Hence $i^*$ is a surjection, 
 for $d \ge 0$, since 
the restriction  $i^*: R_V[z]_d \rightarrow
R_Z[z]_d$ is surjective, and $R_V[z]_d \subseteq
\Gamma(\Delta_V[z],{\cal O}_{\Delta_V[z]}(d))$. Now we prove equality of
multiplicities as in (a). Since, the multiplicity of every $V_\lambda(G)$
in $\Gamma(\Delta_V[z],{\cal O}_{\Delta_V[z]}(d))$ or $R_V[z]_d$ is
now the same, both being equal to the multiplicity of $V_\lambda(K)$
in $\Gamma(\Delta_Z[z],{\cal O}_{\Delta_Z[z]}(d))^*$, it now follows that
$\Gamma(\Delta_V[z],{\cal O}_{\Delta_V[z]}(d)) = R_V[z]_d$, for all 
$d \ge 0$. Hence $R_V[z]$ is integrally closed and 
 $\Delta_V[z]$ is projectively normal (cf. page 126, Hartshorne
\cite{hartshorne}).
\qed

Now let $W$ be any  linear representation of a connected, reductive group
 $K$,
and $R \subseteq K$ a reductive subgroup.
 Fix a point $y \in P(W)$.
Let $Y$ be the smallest  $R$-submodule of $W$ containing  $\hat y$.

\begin{prop} \label{pbranch}
An irreducible $K$-module  $V_\beta(K)$ can occur within 
$R_{W}[y]_d$  only if
an  $R$-module   $V_\alpha(R)$, with $\alpha  \lhd_{R}^K \beta$ occurs within 
$R_{Y}[y]_d$.
 Conversely, if an $R$-module 
$V_\alpha(R)$ occurs in
$R_{Y}[y]_d$  then
there exists a
$K$-module $V_\beta(K)$, with $\beta_{R}^K  \rhd \alpha$,  in  
$R_{W}[y]_d$.
\end{prop} 
\proof 
The embedding $r: Y\rightarrow W$ is $R$-equivariant. 
Hence, we have an $R$-equivariant, closed  embedding 
$r: \check \Delta_Y[y] \rightarrow \check \Delta_W[y]$ of the affine cone
of $\Delta_Y[y]$,
 and the 
corresponding $R$-equivariant  surjection 
$r^*: R_W[y] \rightarrow R_Y[y]$.  Since this surjection is 
degree preserving, by restriction, we get an $R$-equivariant 
surjection $r^*: R_W[y]_d \rightarrow R_Y[y]_d$ for each $d$.

 Let $V_\beta(K)$ be 
any irreducible  $K$-module 
in $R_{W}[y]_d$.
 Arguing as in the Proof of Proposition~\ref{plift}, we can 
conclude that its image under $r^*$ is nontrivial. The image can thus
 be identified 
with an $R$-submodule of $V_\beta(K)$.
If an $R$-module
$V_\alpha(R)$ occurs 
 in this image, then by definition (cf. Section\ref{spartiallystable}), 
 $\alpha \lhd \beta$. 
 Conversely, for every $R$-module 
$V_\alpha(R)$ that appears in $R_Y[y]_d$,
there is a  $K$-module
 $V_\beta(K)$ in $R_{W}[y]_d$ whose image 
contains $V_\alpha(K)$, and hence we must have $\beta \rhd \alpha$.
\qed

\noindent{\bf Proof of Theorem~\ref{tlieoverrefined}:} 
The first and the second statements  follow from Proposition~\ref{plift},
letting $z=v$, $Z=W$.
The third statement  follows from Proposition~\ref{pbranch}.
The fourth   statement follows the statement (a) of 
Theorem~\ref{tlieovernew},
since,  by definition of partial
stability, $v$ is a stable point of $Y$ with respect to the action of $R$.
 \qed

\noindent{\bf Proof of Theorem~\ref{tlieoverpstable}:} 
Suppose $V_\lambda(G)$ occurs in $R_V[v]_d$; i.e., $V_{i_G\lambda}(G)$ 
occurs in $R_V[v]_d^*$. Then by the first statement of 
Theorem~\ref{tlieoverrefined},
$V_{i_G \lambda}(K)$ occurs in $R_{W}[v]_d^*$. That is,
$V_{i_{K} (i_G \lambda)}(K)$ occurs in $R_{W}[v]$. 
It now follows from the third  and fourth
 statements of Theorem~\ref{tlieoverrefined} that $\lambda$ lies over some 
$R_{\hat v}$-admissible weight $\mu$ of $R$. 

Conversely, it  follows from Theorem~\ref{tlieoverrefined} similarly that,
 for every $R_{\hat v}$-admissible dominant weight $\mu$ of $R$,
$R_V[v]$ contains $V_\lambda(G)$ for some
 dominant weight $\lambda$ of $G$ 
lying over $\mu$ at $v$.
 \qed

\section{Application in  complexity theory} \label{scomplexity}
We now specialize the Borel-Weil theorem for partially stable points
(Section~\ref{subspstable}) to the orbit closure problem that 
arises in 
 complexity theory (Section~\ref{sorbitproblem}). 
We  follow the notation of Section~\ref{sorbitproblem}. 
Now $V=\sym^m(Y)$ 
is a linear representation of $G=SL(Y)=SL_l(\C)$,
and $W=\sym^n(X)$ is a representation of $SL(X)=SL_k(\C)$.
 Let $\hat G=GL_l(\C)$.
  Let $i^l$ 
 denote the involution on the weights of $GL_l(\C)$
 so that $V_\lambda(GL_l(\C))^*
=V_{i^l \lambda}(GL_l(\C))$, for a weight $\lambda$.
Recall that every weight $\lambda$  of $GL_l(\C)$ or its dual $i^l(\lambda)$
 corresponds to a
 Young diagram 
of height at most  $l$.
  Every weight of $GL_l(\C)$ that occurs in
 $\C[V]_d^*=\sym^d(V)=\sym^d(\sym^m(Y))$ corresponds to a Young diagram of
size $md$--this will
 be 
 used implicitly in what follows.

\begin{theorem} \label{tlieoverovercomplexity}
(a) Suppose  $g \in P(V)$ is stable with respect to the action of $G$.
Then a Weyl module $V_\lambda(G)$ occurs in $\Delta_V[g]$ iff 
it is $G_{\hat g}$-admissible. 

(b) Suppose $f \in P(V)$ is of the form $\phi(h)$, $h \in P(W)$. 
Then
$V_\lambda(\hat G)$ can occur in $R_V[f]_d$ only if (1) 
 the weight  $i^l(\lambda)$ corresponds to a Young 
diagram with $md$ boxes and height
 at most $k+1$, and (2) $V_{\lambda'}(GL_{k+1}(\C))$, with 
$\lambda'= i^{k+1}\circ i^l(\lambda)$, contains some
$SL_k(\C)_{\hat h}$-admissible module
$V_\mu(SL_k(\C))$, where we consider $SL_k(\C)$ as a subgroup of 
$GL_{k+1}(\C)$ in a natural way.  This means $\mu$ and $\lambda'$ are related
by (a variant of) Pieri's branching rule.

Conversely, for every $SL_k(\C)_{\hat h}$-admissible module
$V_\mu(SL_k(\C))$, there exists a $d$ and $\lambda$ satisfying (1) and 
(2) above such that $V_\lambda(\hat G)$ occurs in  $R_V[g]_d$.
\end{theorem}
\proof (a) follows from Theorem~\ref{tlieovernew} (a). 

(b) The point $f \in P(V)$ is partially stable with defect one
with respect to the action of $\hat G=GL_l(\C)$ on $P(V)$: specifically
$(R,P)$-stable, with $R$ and $P$ as specificed in 
Section~\ref{sorbitproblem}. Now we apply
 Theorem~\ref{tlieoverrefined}
for the action of $\hat G$ on $P(V)$. 
We will only  clarify why the height of $i^l(\lambda)$ is at most $k+1$.
The reductive Levi subgroup of $P$ under consideration is 
 $K\simeq GL_{k+1} \times GL_{l-k-1}$,
 and the subgroup $1 \times GL_{l-k-1}$,
where $1$ denotes the identity in $GL_{k+1}$, is 
contained in the stabilizer $K_{\hat f}$. Suppose $V_\lambda(G)$ 
occurs in $R_V[f]_d$. The irreducible $K$-submodule of $V$  containing 
$f$ is just
$\bar W=\sym^m(\bar X)$ defined in  Section~\ref{sorbitproblem}.
Hence, by Theorem~\ref{tlieoverrefined},
 $V_{i_{K} \circ i^l \lambda}(K)$ is a nonzero $K$-submodule 
of $R_{\bar W}[f]_d$,
where $i_{K}$ is the involution on the weights of $K$.
 By Proposition~\ref{pstableorb},  $V_{i_{K} \circ i^l \lambda}(K)$, and
hence,    $V_{i^l \lambda}(K)$
  must be $K_{\hat f}$-admissible, and hence, 
$1 \times GL_{l-k-1}$-admissible. For any $V_\alpha(GL_l(\C))$,
where $\alpha$ is a Young diagram of height $\le l$,
the $K$-module $V_\alpha(K)$, with the same weight, is equal to
$V_{\alpha_1}(GL_{k+1}) \otimes  V_{\alpha_2}(GL_{l-k-1})$, where 
$\alpha_1$ consists of the first 
 $k+1$ rows of $\alpha$ and
$\alpha_2$ 
consists of the its remaining  $l-k-1$ rows;
here an empty row is treated as a row with zero length.
Let $\alpha=i^l (\lambda)$. Then $V_{\alpha_2}(GL_{l-k-1})$ must be  trivial
since $V_{\alpha}(K)$ is $1 \times GL_{l-k-1}$-admissible: thus $\alpha_2=0$,
 and
 $\alpha_1=\alpha$. It follows that  the length of $\alpha$ is at most $k+1$.
The number of boxes in
$i^l(\lambda)$ must be $md$ since every  irreducible 
$\hat G$-representation occurring in
$\C[V]_d^*=\sym^d(\sym^m(Y))$ has degree $md$.

The rest follows from  Theorem~\ref{tlieoverrefined}
and Theorem~\ref{tlieoverpstable};  details 
 are left to the reader.
 \qed

\section{Representation theoretic data associated with a partially
stable point} 
We extend the definition of the representation theoretic data 
(Definition~\ref{defnnonadmissibledata_new})
to the partially stable case, and illustrate its significance 
with an application to $G/P$. 

\begin{defn} \label{defnnonadmissibledata_new2}
Suppose  $v \in P(V)$ is $(R,P)$ stable, $P=KU$, we say that 
a $G$-submodule $M \subseteq \C[V]_d$ is admissible, with respect to $v$ and
$d$, if 
  $(M^*)^U$ is (1) $(K, \sym^d(W))$-admissible, where $W$ is the 
smallest $K$-submodule of $V$ containing $\hat v$, 
 and (2) it is also $R_{\hat v}$-admissible.
Let $\Sigma_v$ be the set of all nonadmissible 
$G$-submodules of $\oplus_d \C[V]_d = \C[V]$.

Let $\Sigma_v(d) \subseteq \C[V]_d$ be the union of nonadmissible 
$G$-submodules of $\C[V]_d$.
\end{defn}

Basis elements of the $G$-submodules in $\Sigma_v$ will be 
called {\em nonadmissible basis elements}.
The following is a generalization of Proposition~\ref{pcnonadideal_new1}. 

\begin{prop} \label{pcnonadideal_new2}
Suppose  $v$ is $(R,P)$-stable. Then the  $G$-modules in
the representation-theoretic data $\Sigma_v$ associated with $v$ are
contained in $I_V[v]$.
\end{prop} 
\proof 
Let  $P=KU$. Fix any irreducible 
$G$-submodule  $S \subseteq R_V[v]_d$.
The result will follow if we show that every such $S$ 
 is admissible with respect to $v$ and $d$ 
(Definition~\ref{defnnonadmissibledata_new2}). 
 It follows from the first statement of 
Proposition~\ref{peval}  that $S^*$ must contain a $G_{\hat v}$-invariant.
Since $U \subseteq G_{\hat v}$, this implies that  $(S^*)^U$  contains
an $R_{\hat v}$-invariant. 

 Let $W$ be the smallest
$K$-submodule of $V$ containing $\hat v$. It remains to show that
$(S^*)^U$ is $(K, \sym^d(W))$-admissible.
Since $v$, and hence $\hat v$, is stabilized by $U$,
and $U$ is normalized by $K$, $W$ is also a $P$-module with trivial 
$U$-action.
Let $\Phi = G \cdot W \subseteq V$. 
 Consider the induced vector bundle $G \times_P W$ 
(\cite{mumford}) with base space $G/P$ and fibre $W$. Then 
$\Phi$ is  the image of 
 the natural $G$-equivariant map 
$\phi: G \times_P W \rightarrow V$ that maps $(g,x)$, $g \in G$, $x \in W$ to
$g x \in V$. 
We also have the associated   map 
$\tilde \phi:  G \times_P P(W) \rightarrow P(V)$. Since $\tilde \phi$ is
proper, its image $\tilde \Phi$ is closed.   
The $G$-variety  $\Phi$ is
just the affine cone of $\tilde \Phi$, and  is closed. 
$\Delta_V[v]$ is a closed $G$-subvariety of $\tilde \Phi$, and 
its affine cone $\check \Delta_V[v]$ is a closed $G$-subvariety of 
$\Phi$. Hence, $R_V[v]$ is a $G$-summand of the homogeneous 
coordinate ring $R[\tilde \Phi]$ of $\tilde \Phi$. So every 
irreducible $G$ sub-module of $R_V[v]$ can be thought of as 
an irreducible $G$-submodule of $R[\tilde \Phi]$.
An element of $R[\tilde \Phi]_d$ is a regular function on $\Phi$ of 
degree $d$. Its 
 pull back via $\phi$ 
 is a global section of the bundle $B=G \times_P (\sym^d W)^*$. 
Hence, an irreducible $G$-submodule $S$  of $R_V[v]_d$ corresponds to a nonzero
 irreducible $G$-submodule of $\Gamma(G/P, B)$.
 The second statement of 
Proposition~\ref{peval} applied to $B$, in conjunction with Schur's lemma,
implies that, given any such
$S$, $S^*$  must contain a $P$-submodule 
isomorphic to a $P$-submodule of $\sym^d(W)$; i.e., $(S^*)^U$ 
must be $(K,\sym^d(W))$-admissible. 
 \qed

\subsection{Example:  $G/P$} \label{sgmodp}
Proposition~\ref{pcnonadideal_new2}  suggests we study to what extent 
the data
$\Sigma_v$
determines  the ideal $I_V[v]$. 
In this section  we shall show that 
for $G/P$ the data $\Sigma_v$ determines $I_v[v]$ completely.
This observation was a starting point for 
Theorem~\ref{tnonadmissibilitynew} and Conjecture~\ref{conjsftdet}.

Let $G$ be a simply connected, semisimple group $G$ and $P \subseteq G$ its
parabolic subgroup, with Levi decomposition $P=KU$.
 Consider any embedding of $G/P$ in $P(V)$, where 
 $V=V_\lambda(G)$ is an irreducible $G$-representation,
 and $\lambda$ is 
a dominant weight lying in the interior of the face of the dominant Weyl 
chamber in correspondence \cite{fulton}  with  $P$. 
 Let $v \in P(V)$
correspond to its
highest weight vector. Then $G/P$ must actually be the orbit of $v$ in $P(V)$
\cite{fulton}; i.e., 
 $\Delta_V[v]   \simeq G/P$. Recall that $v$ is $(L,P)$-stable, with defect 
zero, where $L$ is the semisimple Levi subgroup of $P$ (Example 1 in
Section~\ref{spartiallystable}).

Basis elements of $\Sigma_v(2)$ 
 are equivalent to 
the Grassman-Pl\"ucker 
syzygies in the case of Grassmanian and, more generally,
the quadratic straightening relations of the standard monomial theory
\cite{smt}
in the ideal of  $G/P$:

\begin{prop} \label{pgmodp}
\begin{enumerate} 
\item $\C[V]_d = V_{d \lambda}(G)^* \oplus \Sigma_v(d)$.
\item $R_V[v]_d= V_{d \lambda}(G)^*$.
\item 
$I_V[v]$ is generated by the basis elements of $\Sigma_v(2)$, the
nonadmissibility data of  degree two.
\end{enumerate}
\end{prop} 

\noindent {\bf Remark:} The second statement is one part
 of the Borel-Weil theorem (cf. Section~\ref{subspstable}).
Compare its proof here with the  one based on Bruhat decomposition
\cite{knapp}.

\proof 
\noindent 1. Since $\C[V]_d^* \simeq \sym^d(V_\lambda)$
 contains a unique highest weight vector with weight 
$d \lambda$, its $G$-module decomposition 
 is of the form 
\begin{equation} \label{eqborel} 
 \C[V]_d^* = V_{d\lambda} + \sum_\mu V_\mu, 
\end{equation}
where each  $\mu$ is some dominant weight smaller than $d \lambda$,
in the usual ordering on the weights. 
Let $W=\C_\lambda$ be the one-dimensional representation (character) of
$P$ corresponding to the weight $\lambda$, so that $\sym^d(W)=\C_{d\lambda}$.
We want to show (cf. Definition~\ref{defnnonadmissibledata_new2})
 that each  $V_\mu
=V_\mu(G)$,  $\mu \not = d \lambda$,  is
not admissible at $v$; i.e., $V_\mu^U$ is not $(K,\C_{d \lambda})$-admissible,
or in other words, that
 $V_\mu$, as a $P$-module, can not  contain 
$\C_{d \lambda}$ as a $P$-submodule (with  trivial $U$-action):
 otherwise let $w \in V_\mu$
 be a basis  vector 
of this one-dimensional module. Since $w$ is invariant under the unipotent 
subgroup of $P$, it  must be the  highest weight vector 
of $V_\mu$, and $\mu$ must belong to the interior of the face of 
the dominant Weyl chamber that corresponds to $P$ \cite{fulton}. 
Moreover,  as a $P$-module, the line $\C w$ corresponding to $w$ 
cannot be isomorphic to $\C_{d \lambda}$ unless 
$\mu = d \lambda$. Hence $V_\mu^* \subseteq \Sigma_v(d)$ 
(Definition~\ref{defnnonadmissibledata_new2}).
 This proves 1.

\noindent 2. By Proposition~\ref{pcnonadideal_new2},
$\Sigma_v(d) \subseteq I_V[v]$, for all $d$.
Hence, this follows from 1. since  $R_V[v]_d$ is clearly nonzero.

\noindent 3. This is now a consequence of
the second fundamental theorem for $G/P$ in 
 the standard monomial theory (cf. Theorem 7.5 in \cite{smt}), which states 
that
 the ideal $I_V[v]$ is 
generated by the functions  in $\C[V]_2$ that vanish on $\Delta_V[v]$.
By 1.
 these are contained in $\Sigma_v(2) \subseteq I_V[v]$.
\qed

\section{SFT for the orbit of a partially stable, excellent  point}
\label{subsidealpstable}
Now we shall prove Theorem~\ref{tnonadmissibilitynew}
for 
partially stable points with defect
zero, by reducing it to the stable case that we have already proved.
Let $V$
be a linear  representation of $G$.  
Let $P \subseteq G$ be a parabolic subgroup with Levi decomposition 
$P=KU=TLU$. We shall assume that the group $R$ in the definition of 
$(R,P)$-stability satisfies the restriction in eq.(\ref{eqrestrictr}),
as it does in our applications (cf. Section~\ref{sorbitproblem}).

A precise statement of Theorem~\ref{tnonadmissibilitynew} in the 
partially stable case is as follows.

\begin{theorem} \label{tpstableorbitnoad}
Let $V=V_\lambda(G)$.
Let $v \in P(V)$ be an $(R,P)$-stable point with defect zero.
Let $W$ be the smallest $K$-submodule  of $V$ 
containing $\hat v$.
Assume that
(1) $L \subseteq R \subseteq K$.
 (2) $R_{\hat v} \subseteq R$ is 
$R$-separable and characterizes $v$,
 considered  as a point in $P(W)$.
Then  the orbit $G v \subseteq P(V)$
 is determined by the representation-theoretic 
 data 
$\Sigma_v$ (Definition~\ref{defnnonadmissibledata_new2}) 
 within some $G$-invariant neighbourhood of the orbit.
Specifically,
there exists a $G$-invariant neighbourhood $Z \subseteq P(V)$
such that $G v$ is a closed subvariety of $Z$ and the zero set (scheme) in
$Z$ of the  basis elements of the $G$-modules in $\Sigma_v$
coincides with $G v$.
\end{theorem} 

For example, suppose  $W=\sym^{n}(X)$ is  embedded via $\phi$ in 
$V=\sym^m(Y)$, as in Section~\ref{sorbitproblem}. Suppose
(1) $f$ is a stable point in $P(W)$ with respect to the action of $R=SL(X)=
SL_{n^2}(\C)$,
(2) $R_{\hat f}$ characterizes $f$ and is $R$-separable.
Then $\phi(f)$ is a partially stable point
of the type above.

Let $\Phi = G \cdot W \subseteq V$ as in the proof of
 Proposition~\ref{pcnonadideal_new2}.
As we observed there, it is the image of 
 the natural $G$-equivariant map 
$\phi: G \times_P W \rightarrow V$ that maps $(g,x)$, $g \in G$, $x \in W$ to
$g x \in V$, and we also have the associated   map 
$\tilde \phi:  G \times_P P(W) \rightarrow P(V)$. Since $\tilde \phi$ is
proper, its image $\tilde \Phi$ is closed.  The $G$-variety  $\Phi$ is
just the affine cone of $\tilde \Phi$.
Let  $R[\tilde \Phi]$ be the  homogeneous coordinate ring of $\tilde \Phi$.

Our goal is to show that the orbit $G v$ of $v$ is determined 
scheme-theoretically by the representation theoretic data within some 
$G$-invariant neighbourhood of the orbit. Since $G v$ is contained in
$\tilde \Phi$, our first goal is to understand the geometry of 
$\tilde \Phi$. Once this is done, we shall be able to reduce the 
present case to the stable case that has already been analyzed.

When $v$ corresponds to  the highest weight vector 
of $V=V_\lambda(G)$, $\tilde \Phi=\Delta_V[v]=G/P$. 
Hence we wish to generalize the results in Section~\ref{sgmodp}.

\subsection*{The geometry of $\tilde \Phi$}

We say that $V_\alpha(G)$ is
$(K,U,W,d)$-admissible if $(V_\alpha(G)^*)^U$
 contains an irreducible $K$-submodule that 
also occurs in 
$\sym^d(W)$, and non-$(K,U,W,d)$-admissible otherwise.

\begin{prop} 
Every $G$-submodule in $R[\tilde \Phi]_d$ is $(K,U,W,d)$-admissible. 
Hence, every non-$(K,U,W,d)$-admissible $G$-submodule of $\C[V]_d$ 
belongs to the homogeneous ideal of $\tilde \Phi$.
\end{prop} 

The proof is an easy modification of that of 
Proposition~\ref{pcnonadideal_new2}.

The following is a generalization of Proposition~\ref{pgmodp}.
Recall that $W$ is a $P$-module with trivial $U$-action
(cf. proof of Proposition~\ref{pcnonadideal_new2}).

\begin{prop} \label{pgmodpextend}
\noindent (1) 
 As a $G$-module,
 $R[\tilde \Phi]_d$
 is isomorphic to the space $\Gamma_d=\Gamma(G/P, \sym^d(W^*))$
of global sections of the vector
bundle $G\times_P \sym^d(W^*)$.

\noindent (2) 
$\C[V]_d= \Gamma_d \oplus (\oplus_\beta V_\beta(G))$, where,
 $V_\beta(G)^*$,  for any $\beta$,  cannot contain an irreducible
 $K$-submodule that also occurs in $\sym^d(W)$. In particular, 
each $V_\beta(G)$  is non-$(K,U,W,d)$-admissible, and hence belongs to 
the ideal of $\tilde \Phi$.

\noindent (3) The ideal of $\tilde \Phi$  is generated (actually 
spanned) by non-$(K,U,W,d)$-admissible $G$-submodules of $\C[V]_d$.

\end{prop} 

For the proof of this proposition we shall need a lemma.
Let $P=KU=TLU$ be the Levi decomposition as above.
We think of the 
root system of $K$ as a subsystem of that of $G$. Let $l$ be any 
linear functional $l$ on the weight space of $G$  with respect to which
the usual ordering of the roots of $G$ is defined; here it is assumed that
$l$
is irrational 
with respect to the weight lattice. 
Let 
\begin{equation} \label{eqknapextend1}
V_\lambda(G)=V_\lambda(K) \oplus \bigoplus_\mu V_\mu(K)
\end{equation} 
 be a
decomposition of $V_\lambda(G)$ as a $K$-module. 
 Let 
$v_\beta$ be the highest weight vector of $V_\beta(K)$ occuring in this
decomposition with respect to $l$. Let $w_T(\beta) = w_T(v_\beta)$
 denote its $T$-weight,
 i.e., the 
weight with respect to the central torus $T \subseteq K$.

The following is a 
 complement to Lemma~\ref{lknapp}.
Let $\phi$ be the projection of 
 the dominant weights of $G$ onto the largest face $F$ of the 
dominant Weyl chamber that is orthogonal (in the Killing norm) to the 
simple roots of ${\cal K}$, the Lie algebra of $K$.
 Note that (1) $w_T(\alpha)=w_T(\phi(\alpha))$,
for any dominant weight, since $w_T(\gamma)=0$ for any simple root $\gamma$ of
$K$,
and (2) $w_T(\phi(\alpha)) \not = w_T(\phi(\beta))$,
if $\phi(\alpha) \not = \phi(\beta)$. Order the projected weights in $F$
according to the restriction   of $l$ to $F$. This induces an order on the 
$T$-weights $w_T(\alpha)$s.

\begin{lemma} \label{lknappextend}
For every $\mu$ in eq.(\ref{eqknapextend1}), $w_T(\mu)=w_T(v_\mu)$ 
 is less than
$w_T(\lambda)=w_T(v_\lambda)$ for an appropriate  $l$.
\end{lemma}
\proof Let $W$ denote the Weyl group of $G$. For a simple root $g$,
let $W_g$ be the reflection in the hyperplane perpendicular 
to $g$.

The weights of $V_\lambda(G)$ are contained in the convex 
hull $C$  of the conjugates of $\lambda$ under the Weyl group elements
\cite{fulton}. Let $A$ be the affine space, perpendicular to $F$, spanned by 
$\lambda$ and
$W_g(\lambda)$'s, where $g$ ranges over the simple roots of ${\cal K}$. 
Its intersection with $C$ 
 is a face  of $C$--call it $L$; it is 
 the smallest face of $C$ containing 
$\lambda$ and
$W_g(\lambda)$, for each simple root  $g$ of  ${\cal K}$. 
\begin{claim} 
The weight vectors of $V_\lambda(G)$, whose weights are contained 
in $L$, span
 the irreducible $K$-submodule $V_\lambda(K) \subseteq V_\lambda(G)$
 with  weight $\lambda$.
\end{claim}
\noindent{\em Proof of the claim:}  
Let ${\cal G}, {\cal K}$ denote the Lie algebras of $G$ and $K$,
 and $U({\cal G}), U({\cal K})$ the corresponding 
universal enveloping algebras.
We know that $V_\lambda(G)$ is spanned by $\alpha v_\lambda$, where 
$v_\lambda$  is the highest weight vector of $V_\lambda(G)$ and 
$\alpha \in U({\cal G})$
 ranges over all monomials in the negative roots of ${\cal G}$.
 If we  order  the roots appropriately,
 the Poincare-Birkoff-Witt theorem implies that $\alpha$ is of the 
form $\alpha_1 \alpha_2$ where $\alpha_2 \in U({\cal K})$ is a monomial 
in the negative roots of ${\cal K}$ and $\alpha_1$ is a monomial in the 
remaining negative roots of ${\cal G}$.
Then $\alpha v_\lambda$ is nonzero with weight in $L$
 iff $\alpha_1=1$. But 
$\alpha_2 v_\lambda$, as $\alpha_2$ ranges over all monomials in 
the negative roots of ${\cal K}$ clearly span $V_\lambda(K)\subseteq 
V_\lambda(G)$. This proves the claim.

It follows from the claim that no
$\mu\not = \lambda$ in eq.(\ref{eqknapextend1}) can belong to $L$.
We shall choose an irrational
$l$ such that the weights of $V_\lambda(G)$ within  $L$
have higher $l$-coordinates than the remaining weights of $V_\lambda(G)$;
it clearly exists.

Consider  the restriction of the
linear function $l$ to $F$. Then  $l(\phi(\alpha))$ is higher than 
$l(\phi(\beta))$ for any weight $\beta$ of $V_\lambda(G)$ not contained 
in $L$. Since
 no $\mu\not = \lambda$ in eq.(\ref{eqknapextend1}) can belong to $L$, 
the result follows.
 \qed

\noindent {\bf Proof of Proposition~\ref{pgmodpextend}:}
Since $w$ is $P$-stable, its stabilizer contains $U$. Since $U$ is 
normalized by $K$, it follows that every point in 
 $W$ is also stabilized by $U$.
By Lemma~\ref{lknapp}, $W= W_\lambda = V_\lambda(G)^U = V_\lambda(K)$. 

The decomposition of $V=V_\lambda$ as a $K$-module is of the 
form:
\[V= V_\lambda = W_\lambda \oplus \bigoplus_\mu W_\mu,\]
where, for  each $\mu$, $w_T(v_\mu) < w_T(v_\lambda)$ 
(Lemma~\ref{lknappextend}).
 Let $W'=\bigoplus_\mu W_\mu$. 
By induction, and using the formula
\[ \C[V]_d^* = \sym^d (V) = \sym^d(W_\lambda \oplus W') 
= \sum_{i+j = d} \sym^i(W_\lambda) \otimes \sym^j(W'),\] 
it follows that $\C[V]_d^*$ has a $K$-module  decomposition of the 
form 
\begin{equation} \label{eqldecompcvd}
 \C[V]_d^* = \sym^d (V) = \sym^d(W) \oplus W_d,
\end{equation}
where the $T$-weight of the highest-weight-vector 
 of each $K$-submodule of $W_d$ is 
strictly smaller than the $T$-weight of the  highest-weight-vector
 of each $K$-submodule in $\sym^d(W)$. Hence
 no irreducible $K$-module
 can occur 
in both $\sym^d(W)$ and 
$W_d$, considered as abstract $K$-modules, i.e.
$\Hom(\sym^d(W), W_d)^K = 0$.

Now consider a 
 $G$-module decomposition 
\begin{equation}   \label{eqdecomp1}
\C[V]_d^* \simeq \sym^d(V_\lambda(G)) = \sum_{\mu} c^\lambda_\mu V_\mu(G),
\end{equation} 
where all $c^\lambda_\mu \ge 0$, and $\mu$ ranges over all dominant 
weights of $G$ less than or equal to $d \lambda$.
We do not know this decomposition explicitly; finding an 
explicit decomposition  is a special 
case of the  unsolved plethysm problem \cite{fulton}.
It follows from eq.(\ref{eqdecomp1}) that
\begin{equation}   \label{equdecompcvd}
\C[V]_d^{*U} \simeq \sym^d(V_\lambda(G))^U = \sum_{\mu} c^\lambda_\mu 
V_\mu(G)^{U} = \sum_{\mu} c^\lambda_\mu 
V_\mu(K), 
\end{equation} 
where the last step follows from  Lemma~\ref{lknapp}. Since $W = V^U$,
  $\sym^d(W)$ is a $U$-submodule of $\sym^d(V)$.
Hence it follows from eq.(\ref{equdecompcvd}) that 
 each weight $\beta$ of $K$ that occurs in $\sym^d(W)$
with nonzero multiplicity $d^\beta$,
also occurs as a weight of $G$ in $\sym^d(V)$ with multiplicity at least 
$d^\beta$. On the other hand, by the Borel-Weil theorem
 and Lemma~\ref{lknapp} (cf. also Frobenius reciprocity \cite{bott}),

\begin{equation} \label{lbwk}
 (\Gamma_d^*)^U = \sym^d(W).
\end{equation}

It follows that  as a $G$-module
\begin{equation} \label{eqdecompcvd}
\C[V]^*_d = \Gamma_d^*  \oplus \bigoplus_\mu c_\mu V_\mu(G),
\end{equation}
for suitable $\mu$'s. 
 On the other hand, comparing this equation with eq.(\ref{eqldecompcvd})
 it follows that no   $V_\mu(G)$ here
 can contain
an irreducible  $K$-submodule that also occurs in
 $\sym^d(W)$.
This proves the second statement of the proposition.

It remains to show that $\Gamma_d$ is a $G$-submodule of $R[\Phi]_d$.
By eq.(\ref{lbwk}), $\Gamma_d^U= \sym^d(W^*)$. Hence by  the
second 
 statement, in conjunction with Lemma~\ref{lknapp},
this is equivalent to showing that $\sym^d(W^*)$ is a $K$-submodule of 
 $R[\Phi]_d^U$. This is clear, since we have the canonical  $U$-equivariant 
embedding of $W$ within $\Phi$, the $U$-action on $W$ being trivial.  \qed

When $\tilde \Phi=G/P$, by the  standard monomial theory, 
we know that nonadmissible basis elements of
degree two generate the ideal of $G/P$ (Section~\ref{sgmodp}).
Analogously, in the context of Proposition~\ref{pgmodpextend},
one can ask for  a degree bound $c$ such that 
the basis elements of  non-$(K,U,W,d)$-admissible $G$-submodules of $\C[V]_d$,
 $d \le c$, generate the ideal of $\tilde \Phi$. This seeks an extension 
of  the standard monomial theory  to
$\tilde \Phi=G W$.

\subsection*{Reduction to the stable case}
Now we are ready to prove Theorem~\ref{tpstableorbitnoad}.
Let $V=V_\lambda(G)$.
 Let $v$ be an $(R,P)$-stable point with
defect zero, as hypothesized, and
$W=V_\lambda(K)$ be the smallest  $K$-submodule of $V$  containing $\hat v$.
 Since 
$L \subseteq R \subseteq K$, $K$ and $R$ are both products of the form
$L T(K)$, $L T(R)$ respectively, where $T(K)$ and $T(R)$ are tori.
Hence an irreducible $K$-module is also an irreducible $R$-module.
In particular,  $W$ is
 an irreducible $R$-module with
 the action of the torus $T(R)$ being determined by a character--i.e.,
the action of $T(R)$ on $P(W)$ is trivial.
Hence, any $R$-invariant subset of $P(W)$ is also $K$-invariant,
and in particular,  $R v = K v \subseteq P(W)$.

 The orbit $G v \subseteq P(V)$ is contained in $\tilde \Phi$.
By Proposition~\ref{pgmodpextend},
the ideal of $\tilde \Phi$  is generated (actually 
spanned) by the non-$(K,U,W,d)$-admissible $G$-submodules of $\C[V]_d$.
These submodules are contained in the nonadmissibility data  $\Sigma_v$ 
associated with $v$ (cf. Definition~\ref{defnnonadmissibledata_new2}).
Let $\hat \Sigma_v$ be the set of remaining $G$-submodules of $\C[V]$ 
in $\Sigma_v$. A $G$-submodule  $M \subseteq \C[V]_d$ belongs to
$\hat \Sigma_v$ iff $(M^*)^U$ 
 is not $R_{\hat v}$-admissible. 
We shall show that there exists a $G$-invariant neighbourhood $Z$ of 
$G v$ in $\tilde \Phi$ such that $G v$ is a closed subvariety of $Z$ and
$G v$ is determined within $Z$ by the data $\hat \Sigma_v$; i.e.,
the zero set of the (basis elements of) the $G$-modules in $\hat \Sigma_v$,
restricted to $Z$, coincides with $G v$ scheme-theoretically.

Consider the $G$-equivariant map 
$\tilde \phi:  G \times_P P(W) \rightarrow \tilde \Phi$. 

\begin{claim} 
 $\tilde \phi^{-1}(v)$ is a point.
\end{claim} 
Proof of the claim: Suppose to the contrary. Then there exists 
$g \not \in P$ and a $w \in P(W)$ such that $\phi(g,w)=v$, i.e., 
$g w = v$, and hence, $w=g^{-1}(v)$. Since $v$ is $(R,P)$-stable,
$U \subseteq G_v \subseteq P$ (Definition~\ref{dpartiallystable}). 
Since $w \in P(W)$, and the $U$-action on $W$ is trivial, 
$$U \subseteq G_w = (G_v)^{g^{-1}} \subseteq P^{g^{-1}}.$$
Thus both $P$ and $P^{g^{-1}}$ contain $U$. This implies that
$P=P^{g^{-1}}$ (by  Lemma 5.2.5 (ii) in \cite{springer}, and 
Corollary 11.17 (iii) in \cite{borel}). Thus $g^{-1}$ normalizes $P$.
Since the normalizer of $P$ is $P$ itself (Theorem 11.16 in \cite{borel}),
it follows that  $g \in P$; a contradiction.

Let us denote the point $\phi^{-1}(v)$ by $\tilde v$. 
 Since $\tilde \phi$  is
surjective, to show that 
$G v$ is  scheme-theoretically determined within a $G$-invariant neighbourhood 
 by the data $\hat \Sigma_v$, it  suffices to 
show that $\tilde \phi^{-1}(G v) = G \cdot \tilde \phi^{-1} (v) = G \tilde v 
\subseteq G \times_P P(W)$
 is determined scheme-theoretically within
some $G$-invariant neighbourhood by the set $\tilde \phi^{-1}(\hat \Sigma_v)$
of the pull backs of the 
$G$-modules in $\hat \Sigma_v$. But since $G_{\tilde v} =G_v \subseteq P$,
the normal space to $G \tilde v$ can be identified with the normal space 
to its restriction to the slice $\tilde \phi^{-1}(P(W)) \simeq P(W)$, which 
in turn, corresponds to 
 the normal space to the orbit $R v = K v \subseteq P(W)$.
By the Jacobian criterion (Proposition~\ref{galbal}), it now 
suffices to show that
the set $\tilde  \phi^{-1}(\hat \Sigma_v)_{P(W)}$ of 
the restrictions of the  modules in $\tilde  \phi^{-1}(\hat \Sigma_v)$
  to the  fixed  slice 
$\tilde \phi^{-1}(P(W)) \simeq P(W)$ of 
the bundle  $G \times_P(W)$ determines the orbit of $R v = K v \subseteq P(W)$
within some $K$-invariant neighbourhood of
 this orbit.

 By Proposition~\ref{pgmodpextend}, $R[\Phi]_d$ is isomorphic to 
the space $\Gamma_d= \Gamma(G/P,\sym^d(W^*))$ of global sections of the 
bundle $G \times_P \sym^d(W^*)$. By the Borel-Weil theorem  and 
 Lemma~\ref{lknapp} (see also the Frobenius reciprocity in \cite{bott}), the 
set of restrictions of the modules in $\Gamma_d$ to the slice $P(W)$ 
can be identified with
 $\Gamma_d^U$: if $M \in R[\Phi]$ and 
is isomorphic to $V_\lambda(G)$, then the restriction of $\tilde \phi^{-1}(M)$
to $P(W)$ corresponds to $M^U$, which is isomorphic to 
$V_\lambda(K)$ (Lemma~\ref{lknapp}). 
 Hence, the restrictions of the  modules in $\tilde \phi^{-1}(\hat \Sigma_v)$
 to the slice $P(W)$   consists of precisely  the $K$-modules in
$\C[W]$ that do not contain any $R_{\hat v}$-invariant. Since 
$K$ and $R$ are of the form
$L T(K)$, $L T(R)$,
 an irreducible $K$-module is also an irreducible $R$-module, and
 the subspace of $\C[W]$ spanned by  non-$R_{\hat v}$-admisible
 $K$-submodules coincides with the subspace spanned by 
 non-$R_{\hat v}$-admisible $R$-submodules. Thus, 
 $\tilde \phi^{-1}(\hat \Sigma_v)_{P(W)}$
 consists of precisely  the
 non-$R_{\hat v}$-admisible $R$-modules in
$\C[W]$.
Since $v \in P(W)$ is stable with respect to the action of $R$ on $P(W)$,
we can now apply  Theorem~\ref{tnonadmissibility_new2}
 for the stable case.
 It implies 
 that $R v \subseteq P(W)$ has an $R$-invariant, and hence, $K$-invariant,
 neighbourhood $Y$ such that $R v$ as a subvariety 
of $Y$ is determined scheme-theoretically by  
$\tilde \phi^{-1}(\hat \Sigma_v)_{P(W)}$.

This shows that
$\tilde  \phi^{-1}(\hat \Sigma_v)_{P(W)}$ 
 determines the orbit of $R v = K v \subseteq P(W)$
within a $K$-invariant neighbourhood of
 the orbit.

This proves Theorem~\ref{tpstableorbitnoad}. \qed 

\section{$G$-separability} \label{sseparable}
We now study the notion of $G$-separability (Definition~\ref{dgseparable}),
 which is of interest in the 
context of Theorem~\ref{tnonadmissibilitynew}.

\begin{prop} \label{pbasicsep}
\begin{enumerate}
\item A semisimple group $H$, embedded in $G=H \times H$ diagonally,
is strongly $G$-separable.
\item $H=SL_k(\C)$ is a strongly $G$-separable subgroup of
 $G=SL_n(\C)$ if $k > (n+1)/2$.
\item 
  $H=SL_k(\C) \times SL_l(\C) \subseteq 
G=SL_{k+l}(\C)$, with  natural embedding, is strongly $G$-separable.
\end{enumerate}
\end{prop}

\noindent{\bf Remark:} The last statement 
 can be generalized to semisimple Levi subgroups
of maximal parabolic subgroups of classical simple groups, if one
uses, instead of the decomposition formula in eq.(\ref{eqfulton1}) below,
Littelmann's restriction rule \cite{littelmann}.

\proof
\noindent (1) By Schur's lemma, a
 $G$-module $V_\alpha(H) \otimes V_\beta(H)$,
where $\otimes$ denotes the external tensor product here, is $H$-admissible
iff $V_\beta(H) \simeq V_\alpha(H)^*$, i.e. $\beta=i_H(\alpha)$,
where $i_H$ is the involution on dominant $H$-weights
 (Section~\ref{spartiallystable}).
Any nontrivial representation $V_\lambda(H)$ occurs in the non-$H$-admissible
$G$-module $V_\lambda(H) \otimes 1_H$, where $1_H$ denotes the trivial
$H$-module.
So $H$ is clearly $G$-separable. 

 Strong $G$-separation follows from the following more general fact.
\begin{claim} 
 $V_\lambda(H)$ occurs in the non-$H$-admissible
$G$-module $V_{\delta} (H) \otimes V_{\rho}(H)$, $\delta=\lambda + \beta$
$\rho=i_H (\beta)$, for 
any dominant $H$-weight $\beta$.
\end{claim}
Proof of the claim: By Schur's lemma,  this is equivalent to showing that 
$$
\begin{array} {lcl}
\Hom(V_\delta(H) \otimes V_\rho(H), V_\lambda(H)) &= &
V_{\lambda+\beta}(H)^* \otimes V_\rho(H)^* \otimes V_\lambda(H) \\
& = & V_{\lambda+\beta}(H)^* \otimes V_\beta(H) \otimes V_\lambda(H)
\end{array}$$
contains an $H$-invariant. By Schur's lemma again, this 
is equivalent to showing that $V_{\lambda+\beta}(H)$ occurs in
$V_\beta(H) \otimes V_\lambda(H)$, which is clear.

\ignore{ This readily follows from 
Littelmann's rule \cite{littelmann}  for decomposing the tensor product
$V_\delta(H) \otimes V_\rho(H)$ as an $H$-module.
Let $\tilde \delta$ and $\tilde \rho$ be any piecewise linear paths in
the weight space of $H$ connecting $0$ with $\delta$ and $\rho$,
respectively, and lying completely within the dominant Weyl chamber of 
$H$. Let $\tilde \delta * \tilde \rho$ be its concatenation, which ends at
$\delta + \rho$. Let 
$B=B_{\tilde \rho}$ be the   set of 
 piecewise linear paths  obtained from 
  $\tilde \rho$ by applying (any combination) of 
Littelmann's root operators. Then
\[ V_\delta(H) \otimes V_\rho(H) = \oplus_{\pi(1)} V_{\pi(1)}(H),\]
where $\pi$ ranges over all paths in the dominant Weyl chamber of 
$H$ of the form $\tilde \delta * \eta$
 for some path $\eta \in B_{\tilde \rho}$.
Now let $\tilde \beta$, $\tilde \rho$,  and $\tilde \lambda$ 
 be the straight paths connecting $0$ with $\beta$, $\rho$, and $\lambda$,
respectively. By Schur's lemma,
 $V_\beta(H) \otimes V_\rho(H)$ contains the trivial
$H$-module. Hence by the decomposition formula above,
 there exists some path $\pi \in B_{\tilde \rho}$
such that $\tilde \beta * \pi$ lies in the dominant Weyl chamber and 
ends at $0$.
Let $\tilde \delta = \tilde \lambda * \tilde \beta$. 
Then $\tilde \delta * \pi$ also lies in the dominant Weyl chamber and 
it ends at $\lambda$. Hence, by the decomposition formula,
$V_\lambda(H)$ occurs in  $V_{\delta} (H) \otimes V_{\rho}(H)$. 
This proves the claim. }

\noindent (2) Consider a nontrivial $V_\lambda(SL_k(\C))$, where $\lambda$ is
a Young diagram of height $h$ less than
$k$. We shall exhibit a non-$H$-admissible $V_\mu(SL_n(\C))$ containing it.
If $h$ is greater than $n-k$, we let $\mu = \lambda$.
Otherwise, let $\mu$ be a Young diagram obtained by adding 
$n-k-h+1$ boxes  to the first column of $\lambda$. Its height is $n-k+1 < k$.
By Pieri's branching rule, it is easy to see  that 
$V_{\mu}(SL_n(\C))$ contains $V_\lambda(SL_k(C))$ but not the trivial 
representation of $SL_k(\C)$.
More generally, if $\mu'$ is a Young diagram obtained by appropriately
 extending, i.e., adding boxes to the   first $n-k$ rows of $\mu$, then
$V_{\mu'}(SL_n(\C))$ contains $V_\lambda(SL_k(C))$ but not the trivial 
representation of $SL_k(\C)$.
There are infinitely many such $\mu'$s. So $SL_k(\C)$ is strongly 
separable.

\noindent (3) 
Assume that $k \ge l$, the other case being similar.
Consider a nontrivial  $H$-module
 $L=V_\alpha(SL_k(\C)) \otimes V_\beta(SL_l(\C))$, where
$\alpha$ and $\beta$ 
 correspond to Young diagrams of height less than $k$ and $l$ 
respectively. We shall exhibit a non-$H$-admissible $G$-module $V_\lambda(G)$
containing it. 
We identify $\alpha$ and $\beta$  with the 
partitions:
$\alpha=(\alpha_1,\alpha_2,\ldots)$, where $\alpha_i$ denotes the length of 
the $i$th row of the corresponding Young diagram, and 
$\beta=(\beta_1,\beta_2,\ldots)$.
We proceed by cases.

\noindent {\bf Case 1:} Either $\alpha$ does not correspond to a rectangular 
Young diagram of height $l$, or $\beta$ is not trivial.

 Let $\lambda=\alpha+\beta=
(\alpha_1+ \beta_1, \ldots)$. Note that the height of $\lambda$ 
is less than $k$.
We have \cite{fulton},
\begin{equation} \label{eqfulton1}
V_\lambda(GL_{l+k}(\C))=\sum_{\rho,\delta} N^\lambda_{\rho,\delta}
V_\rho(GL_{k}(\C)) \otimes V_\delta(GL_l(\C)),
\end{equation}
where $N^\lambda_{\rho,\delta}$ denotes the Littlewood-Richardson 
coefficient. From this it easily follows that 
$V_\lambda(SL_{k+l}(\C))$ contains the representation 
$V_\alpha(SL_k(\C)) \otimes V_\beta(SL_l(\C))$ of 
$SL_k(\C) \times SL_l(\C)$. But it cannot contain
 the  trivial $H$-representation:
 If $\rho\not = 0$ and $\delta$ (possibly zero)
 correspond to  rectangular 
Young diagrams with height $k$ and $l$ respectively--so that 
$V_\rho(SL_k(\C))$ and  $V_\delta(SL_l(\C))$ are trivial--then 
$N_{\rho,\delta}^\lambda$ is easily seen to be zero; otherwise
the height of $\lambda$ will be at least $k$. On the other hand,
if $\rho=0$, then $\lambda=\delta$. 
Since the height of $\beta$ is less than $l$,
the definition of $\lambda$ then implies that $\alpha=\delta$ and $\beta=0$;
 a contradiction.

More generally,  let $\alpha'$ be any Young diagram obtained from
$\alpha$ by adding columns of length $k$.
Let $\lambda'=\alpha'+\beta$.
Then $V_{\lambda'}(SL_{k+l}(\C))$ also contains 
$V_\alpha(SL_k(\C)) \otimes V_\beta(SL_l(\C))$ but not the 
trivial representation of $SL_k(\C) \times SL_l(\C)$. 
Moreover, there are infinitely many such $\lambda'$s.

\noindent {\bf Case 2:}  $\alpha$ is rectangular of height $l$
and width $w$,
and $\beta=0$.

We can assume that $k>l$; otherwise $V_\alpha(SL_k(\C))$ too will be trivial.
 For any integer $r \ge 0$, let $\lambda$ be the Young 
diagram whose first $r$ columns are of height $k$, the $(r+1)$-st column 
is of length $l+1$, the columns numbered $r+2,\ldots, r+w$ are of height
$l$, and the column numbered $r+w+1$ is of height $l-1$. Then 
it follows from eq.(\ref{eqfulton1}) that 
$V_\lambda(GL_{l+k}(\C))$ contains 
$V_\rho(GL_k(\C)) \otimes V_\delta(GL_l(\C))$, where $\rho$ is obtained 
from $\alpha$ by adding to its left $r$ columns of length $k$,
 and $\delta$ consists of
a single column of  height $l$.
Clearly  $V_\rho(GL_k(\C)) \otimes V_\delta(GL_l(\C))$
is isomorphic to $V_\alpha(SL_k(\C)) \otimes V_\beta(SL_l(\C))$ as 
an $SL_k(\C) \times SL_l(\C)$-module. But it does not contain 
the trivial $SL_k(\C) \times SL_l(\C)$-module; this too follows from
 eq.(\ref{eqfulton1}). Moreover, there are infinitely many such $\lambda$.

 This proves 
strong $G$-separability of $H$.
  \qed

\ignore{
\noindent (3) This generalizes the second statement of (2). Accordingly,
we will generalize the idea in the proof of (2) above,
using  the path model of the standard monomial theory
\cite{smt}.
Let $K$ be a reductive Levi subgroup of a parabolic subgroup of $G$.
Let ${\cal K}$ and ${\cal G}$ denote their Lie algebras.
A root system of ${\cal K}$ is a subsystem of that of ${\cal G}$.
Let $V_\lambda(G)$ be any irreducible $G$-module. 
Let 
$B_\lambda$ be the set of Lakshmibai-Seshadri paths that index a 
standard basis of $V_\lambda(G)$.

By Littelmann's restriction rule \cite{littelmann},
 $V_\lambda(G)$ 
decomposes as a $K$-module into the direct sum 
$\oplus_{\eta} V_{\eta(1)}(K)$, where $\eta$ runs over all 
Lakshmibai-Seshadri paths  in $B_\lambda$ 
contained in the dominant Weyl chamber of the root system of ${\cal K}$.

Now let $P$ be a  maximal parabolic subgroup $P$ of $G$,
with Levi decomposition $P=KU=TLU$. We shall prove $G$-separability of $L$;
the proof for $K$ is similar, and simpler.
Let $\Pi_P$ be the subset of simple roots of ${\cal G}$ that correspond to
$P$; it consists of all but one of the simple roots of ${\cal G}$.
Let $\lambda$ be a dominant weight of $L$. Then there is a unique weight
in the dominant Weyl chamber of $K$, which restricts to $\lambda$ on $L$
and to the trivial weight on the torus $T$; we shall denote this dominant 
weight of $K$ by $\lambda$ again. 
Let $W_G$ be the Weyl group of $G$.
Let $\hat \lambda$ be the unique weight in the $W_G$-orbit of $\lambda$ 
that is contained in the 
dominant Weyl chamber of ${\cal G}$. 

\begin{claim} 
$V_{\hat \lambda}(G)$ contains $V_{\lambda}(K)=V_\lambda(L)$
 but not the trivial
representation of $L$.
\end{claim}

Proof of the claim: That $V_{\hat \lambda}(G)$ contains $V_{\lambda}(K)$
follows from Littelmann's restriction. That it does not contain the 
trivial representation of $L$ follows 
by case analysis of the root system.
 We have already shown this for $A_n$ in (2). Since all proofs are similar,
we will  consider just one more case $D_n$. 
Let $e_1-e_2, e_2-e_3, \cdots, e_{n-1}-e_n,
e_{n-1} + e_n$ be the ordered set of
 simple positive roots, labeling the nodes of the 
Dynkin diagram.
 Let $P$ be a maximal parabolic subgroup that corresponds to the
Dynkin diagram obtained by removing the node labeled by $L_j-L_{j+1}$, 
$j<n-1$; the remaining cases are similar.
Let $L=L_1\times L_2$, where $L_1\simeq SL_{j-1}(\C)$.
 Suppose for some Lakshmibai-Seshadri path
$\eta$, as specified in Littelmann's restriction rule, the weight $\eta(1)$,
when restricted to $L$ is trivial. Then, it is easy to see that 
 $\eta(1)$
 must be a multiple of $e_1+ \cdots + e_j$. This is not possible by 
a version of Frobenius reciprocity
 (to be elaborated). The proof for the remaining root systems is 
similar. This proves the claim.

\noindent (4) Given a representation $V_\mu(H)$, the representation 
$V_0(H) \hat \otimes V_\mu(H)$ of $H \times H$, where $\hat \otimes$ 
denotes the external tensor product, contains, as an $H$-module,
 $V_\mu(H)$ but not the trivial representation of $H$. 
\qed 

\noindent {\em Remark:} Check if the previous subgroups are also strongly 
separable. 
}

For us, it is important  to know if 
the stabilizers of the points that arise in the context of complexity
theory are separable (cf. Section~\ref{sorbitproblem}).
The connected component 
of the stabilizer of $\det(Y)$ in $SL_{n^2}(\C)$,
where $Y$ is an $n\times n$-matrix, contains 
 $SL_n(\C) \times SL_n(\C) \subseteq SL(Y)=
SL_{n^2}(\C)$ (Section~\ref{subspermvsdt}).
 Regarding this subgroup we make the following:

\begin{conj}
$SL_n(\C) \times SL_n(C)$ is a strongly separable subgroup of $SL_{n^2}(\C)$.
\end{conj}
Here the embedding corresponds to the natural embedding $SL(V) \otimes SL(V) 
\subseteq SL(V \otimes V)$, $V=C^n$.
Specifically, letting $V_\lambda(n)$ denote  $V_\lambda(SL_n(\C))$ in what 
follows, the conjecture can be reformulated as follows:

\begin{conj} \label{conjsln}
For  every nontrivial Weyl module $V_\lambda(n) \otimes V_\mu(n)$ 
of $SL_n(\C) \times SL_n(\C)$,
such that $|\lambda| = |\mu|  \ (\modulo n)$, there exist (infinitely many)
 Weyl modules 
$V_\rho (n^2)$ of $SL_{n^2}(\C)$ whose decomposition as an
$SL_n(\C) \times SL_n(\C)$-module contains $V_\lambda(n) \otimes V_\mu(n)$
but not the trivial $SL_n(\C) \times SL_n(\C)$-module. 
\end{conj} 

 The restriction $|\lambda| = |\mu| \ (\modulo n)$ is required to
ensure  (cf. Definition~\ref{dgseparable}) that 
$V_\lambda(n) \otimes V_\mu(n)$ occurs in some 
representation of $SL_{n^2}(\C)$;
cf. eq.(\ref{eqglmdecomp}) below.

The conjecture  can be reformulated in terms of the symmetric group as follows.
Let $\hat V_\gamma(n^2)$ be a Weyl module  of $GL_{n^2}(\C)$. 
Embed $GL_n(\C) \times GL_n(\C)= GL(\C^n) \times GL(\C^n)$ in
$GL(\C^n \otimes \C^n)= GL_{n^2}(\C)$. The decomposition of  
$\hat V_\gamma(n^2)$
 as a $GL_n(\C)\times GL_n(\C)$-module is of the form:

\begin{equation} \label{eqglmdecomp}
 \hat V_\gamma(n^2) = \sum_{\alpha,\beta}
c_{\alpha,\beta,\gamma}  \hat V_\alpha(n) \otimes \hat V_\beta(n);
\end{equation}
here $c_{\alpha,\beta,\gamma}$ can be nonzero only if 
$|\alpha| = |\beta|=|\gamma|$.
To get  the decomposition of 
$\hat V_\gamma$ as an $SL_n(\C) \times SL_n(\C)$ module, 
we reduce the Young diagrams occurring on the right hand side by removing 
columns of length $n$. This does not change their sizes modulo $n$; this 
explains  the restriction $|\lambda|=|\mu| \ (\modulo n)$ in
the conjecture.
By Littlewood's symmetry conditions
 (\cite{fulton}),
 the coefficients  $c_{\alpha,\beta,\gamma}$ do not depend on the ordering of
$\alpha,\beta$ and $\gamma$

Given a   Young diagram $\delta$, $|\delta|=m$, 
let $W_\delta$  denote the corresponding 
irreducible representation, the Specht module, of the symmetric group $S_m$.
Then the  coefficient $c_{\alpha,\beta, \gamma}$ occurring 
in the preceding 
  decomposition is the same as the one occurring in the decomposition 
of the tensor product $W_\alpha \otimes W_\beta$
 as an $S_m$-module,
\[ W_\alpha \otimes W_\beta = \sum_\gamma c_{\alpha,\beta,\gamma} W_\gamma,\]
 where $m=|\alpha|=|\beta|$; cf. \cite{fulton}.

For any $\lambda$ of height less than $n$, and $m=|\lambda| \ \ (\modulo n)$, 
let $\lambda(m)$ be the unique Young diagram of size $m$ 
obtained by adding to $\alpha$  columns of length $n$. Then the 
preceding conjecture is equivalent to saying that: 

\noindent 
For every nontrivial 
pair of Young diagrams $(\lambda, \mu)$ of height less than $n$,
and such that $|\lambda| = |\mu| \ (\modulo n)$, there exists 
an $m=|\lambda|=|\mu| \ (\modulo n)$, $m \ge n$,
 and a $\rho$ of size $m$ such that 
$W_\rho$ occurs in the decomposition of $W_{\lambda(m)} \otimes W_{\mu(m)}$ 
as an $S_m$-module, but not in the decomposition of 
$W_\delta \otimes W_\delta$, where $\delta$ is the rectangular Young 
diagram of height $n$ and size $m$.

If  $|\lambda| = |\mu| \not = 0 \ (\modulo n)$,
the last 
restriction is vacuous, because 
no such $\delta$ exists, and hence:
\begin{prop} 
If  $|\lambda| = |\mu| \not = 0 \ (\modulo n)$, Conjecture~\ref{conjsln} holds.
\end{prop}

 So, let us assume 
that $|\lambda| = |\mu|  = 0 \ (\modulo  n)$ in what follows.

\begin{prop} \label{psl2}
Conjecture~\ref{conjsln} holds for $n=2$.
\end{prop}

The main difficulty in extending the proof below to $n>2$ is that
an explicit decomposition of the tensor product of two arbitrary 
Specht modules is not yet known.

\proof 
We need to show that 
for every nontrivial pair of  $(\lambda, \mu)$ of row-shaped Young diagrams,
with $|\lambda|$ and $|\mu|$ even, there exists  an even $m$
 and a $\rho$ of size $m$ such that 
$W_\rho$ occurs in the decomposition of $W_{\lambda(m)} \otimes W_{\mu(m)}$ 
as an $S_m$-module, but not in the decomposition of 
$W_\delta \otimes W_\delta$, where $\delta$ is the rectangular Young 
diagram of height $2$ and width $m/2$.  We shall show that that there 
exist such a $\rho$ for every large enough $m \ge 4 (|\lambda| + |\mu|)$.
Fix such an $m$.

Given a Young diagram $\gamma$, we shall let $\gamma_i$ denote the 
number of boxes in its $i$th row from the top. We assume that 
the topmost row has the highest length in the diagram.
We shall denote $\lambda(m)$ and $\mu(m)$ by 
$\bar \lambda$  and $\bar \mu$  respectively. Since $\lambda$ and
$\mu$ are row shaped, we shall let $\lambda$ and $\mu$ denote the lengths
of their row as well.
Since $|\bar \lambda|=|\bar \mu|=m$,
$\bar \lambda_2 - \bar \lambda_1 = \lambda$ and 
$\bar \mu_2 - \bar \mu_1 = \mu$, we have $\bar \lambda_2= m/2 - \lambda/2$
and $\bar \mu_2 = m/2 - \mu/2$.
Since $\bar \lambda$, $\bar \mu$ and $\delta$ have two rows,
 we can use the  decomposition formula of Remmel and Whitehead \cite{remmel}.

First, we shall try to find a required $\rho$ with two rows.
Let $(a,b)$, $a \ge b$, denote the two-rowed Young diagram with the top row 
of length $a$ and the bottom row of length $b$. 
Suppose we are given 
Young diagrams $(k,h), (r,l), (d,c)$ of size $m$.
Because of Littlewood's symmetry
conditions we can assume that  $l \le h \le c$. With this condition,
The  formula in 
\cite{remmel} (Thm 3.3)
says that 
\begin{equation} \label{eqremwhite}
  c_{(r,l),(k,h),(d,c)} = (1+ w - v) \chi(w \ge v),
\end{equation}
where  $w = \lfloor (l + h -c)/2 \rfloor$,  
$v=\mbox{max}(0,\lceil (l+h+c-m)/2 \rceil)$, and the function $\chi$ is 
one if $w \ge v$ and zero otherwise.

By Littlewood's symmetry condition, 
$c_{\delta,\delta,\rho}=c_{\rho,\delta,\delta}$. Applying 
the preceding  formula with $(r,l)=\rho$, and $(k,h)=(d,c)=\delta=(m/2,m/2)$,
we conclude that this coefficient  is nonzero iff 
$\lfloor \rho_2/2 \rfloor  \ge \lceil \rho_2/2 \rceil$. That is, iff 
$\rho_2$ is even. So we need to find  a $\rho$, with $\rho_2$ odd, such 
that $c_{\bar \lambda, \bar \mu,\rho}$ is nonzero. 
Because of symmetry, we can assume that 
$\bar \lambda_2 \le \bar \mu_2$.
We will try to find $\rho$ such that 
\begin{equation} \label{eqremmel1}
\rho_2 \le \bar \lambda_2.
\end{equation} Then 
setting  $(k,h)=\rho$, $(r,l)=\bar \lambda$,
and $(d,c)=\bar \mu$ in eq.(\ref{eqremwhite}), we conclude that
$c_{\bar \lambda, \bar \mu,\rho} = c_{\rho,\bar \lambda, \bar \mu}$ 
is nonzero 
iff 
\begin{equation}
\lfloor (\rho_2 + \bar \lambda_2 - \bar \mu_2)/2 \rfloor \ge 
\mbox{max}(0, \lceil (\rho_2 + \bar \lambda_2 + \bar \mu_2 -m)/2 \rceil),
\end{equation} 
i.e., iff
\begin{equation}  \label{eqrem} 
\lfloor (\rho_2 -  \lambda/2 +  \mu/2)/2 \rfloor \ge 
\mbox{max}(0, \lceil (\rho_2 -  \lambda/2 -  \mu/2)/2 \rceil).
\end{equation}
We now proceed by cases.

\noindent Case 1: $\mu \not = 0$, 

In this case the condition in eq.(\ref{eqrem}) 
 can be satisfied if 
\begin{equation} \label{eqremmel2}
\rho_2 \ge (\lambda + \mu)/2,
\end{equation}
and $\mu \ge 2$, which holds since $\mu$ is nonzero and even. 
But there are many odd $\rho_2$'s such that eq(\ref{eqremmel1})
and eq(\ref{eqremmel2}) are satisfied if, say,
$m\ge 4 (\lambda + \mu)$.

\noindent Case 2: $\mu = 0$, and $\lambda/2$ is odd.

In this case,  eq.(\ref{eqrem}) is satisfied if 
 we let  $\rho_2 = \lambda/2$, which is nonzero--otherwise
$(\lambda,\mu)$ will be trivial--and odd, as required.
Since $m$ is large enough, eq.(\ref{eqremmel1}) is also satisfied.

It remains to consider

\noindent Case 3: $\mu=0$ and $\lambda/2$ is even.

In this case, the required two-rowed $\rho$ does not exist. So 
 we shall find an appropriate  $\rho = (\rho_1,\rho_2,
\rho_3,\rho_4)$ with four rows such that $\rho_3=\rho_4$.

Given Young diagrams $(k,h)$, $(m,l)$, $(d,c,a,a)$ (entries in the 
nonincreasing order) with $m$ boxes 
such that $a>0$ and $\lceil (h+1)/2 \rceil \le h-c$,
the  Remmel-Whitehead  formula (\cite{remmel}, Theorem 3.1) says that

\begin{equation} \label{eqremmelfour}
c_{(k,h),(m,l),(d,c,a,a)} = 
\sum_{r=h-c}^{\min(l,\lfloor \f {l-a+h-c} 2 \rfloor)} 1 -
\sum_{r=\max(a,l+h+a-m-1)}^
{\min(l,\lfloor \f {h-1}2 \rfloor, \lfloor \f {l+h+a+c-m-1}2 \rfloor)} 1.
\end{equation} 

We will set
 $(k,h)=(m,l)=\delta = (m/2,m/2)$ and $(d,c,a,a)= \rho=(\rho_1,\rho_2,
\rho_3,\rho_4)$ in this formula. For the formula to be applicable, 
 we need to ensure that 
\begin{equation}  \label{eqrho1}
\lceil \lceil (h+1)/2\rceil = (m/2 +1) /2 \rceil \le h-c= m/2 - \rho_2.
\end{equation} 
If, in addition,
\begin{equation} \label{eqrho2}
\rho_2 + \rho_3 < m/2,
\end{equation}
we get that
\[
\begin{array}{ll}
c_{\delta,\delta,\rho} & 
  =  \sum_{m/2-\rho_2}^{\lfloor (m-\rho_3-\rho_2)/2 \rfloor}  1 -
\sum_{r=\rho_3}^{\lfloor \f {\rho_3 + \rho_2 -1} 2 \rfloor} 1 \\
& = \lfloor (m-\rho_3-\rho_2)/2) \rfloor - (m/2-\rho_2) + \rho_3 - 
\lfloor \f {\rho_3 +\rho_2 -1} 2 \rfloor, \\
& = \lfloor \f {\rho_2 -\rho_3} 2 \rfloor - 
   \lfloor \f {\rho_2 -\rho_3 - 1} 2 \rfloor,
\end{array}
\]
which is $1$ if $\rho_2 - \rho_3$ is even, and zero otherwise. 

So we need to find a $\rho$ with $\rho_2 -\rho_3$ odd, satisfying 
eq.(\ref{eqrho1}) and (\ref{eqrho2}),  such that 
$c_{\bar \mu,\bar \lambda,\rho}$ is nonzero.
Since $\mu=0$, we have $\bar \mu_1=\bar \mu_2=m/2$. Also, recall that
$\bar \lambda_2= m/2 -\lambda/2$.  
Set
$(k,h)=\bar \mu = (m/2,m/2)$, $(m,l)=\bar \lambda$ and 
$(d,c,a,a)=\rho$ in eq.(\ref{eqremmelfour}). 
We shall choose a four rowed $\rho$, with nonzero  $\rho_3$,  such that 
$\rho_2 -\rho_3$ is odd,  $\rho_2$ and $\rho_3$ are sufficiently 
larger than $\lambda$, and also 
such that the difference between $m/2$ and $\rho_2 + \rho_3$ is sufficiently
larger than $\lambda$. This 
is possible if $m$ is large enough compared to $\lambda$.
  In this case, the upper index of the first sum in
eq.(\ref{eqremmelfour}) becomes 
$\lfloor \f {\bar \lambda_2 -\rho_3 + m/2 -\rho_2}2\rfloor$, and 
the lower index is $m/2-\rho_2$. So the contribution of the first term is 
$\lfloor \f {\rho_2 - \rho_3 - \lambda/2} 2 \rfloor$.
Since $\lambda$ is nonzero, the lower index of the second sum in
eq.(\ref{eqremmelfour}) is equal to $\f \lambda 2 - 1 + \rho_3$. 
The 
upper index, assuming that $m$ is large enough, and $m/2- \rho_2 - \rho_3$ 
is sufficiently larger than $\lambda$, becomes 
$\lfloor \f {-\lambda/2 + \rho_2 + \rho_3 -1} 2 \rfloor$. Assuming that 
$\rho_2$ and $\rho_3$ are sufficiently larger than $\lambda$, it is 
larger than the lower index.
 Hence the  second 
term becomes 
\[ \f \lambda 2 - 1 + \rho_3 - 
\lfloor \f {-\lambda/2 + \rho_2 + \rho_3 -1} 2 \rfloor = 
\lambda/2 -1 + \lfloor \f {\lambda/2 - \rho_2 + \rho_3 + 1} 2 \rfloor.\]
Thus
\[ c_{\bar \mu,\bar \lambda,\rho} = 
\lfloor \f {\rho_2 - \rho_3 - \lambda/2} 2 \rfloor + 
\lambda/2 -1 + \lfloor \f {\lambda/2 - \rho_2 + \rho_3 + 1} 2 \rfloor
 = \lambda/2 -1.\]
This is
nonzero, since $\lambda/2$, being  nonzero and even, is at least two.
So we can choose a $\rho$, with $\rho_2-\rho_3$ odd, and subject to the
preceding conditions, as required.
\qed

\end{document}